\def\nk{n_{\rm b}}

\def\Pb{P_{\rm b}}

\def\rfr#1{Equation\,(\ref{#1})}
\def\rfrs#1#2{Equations\,(\ref{#1})--(\ref{#2})}
\def\Rfr#1{Equation\,(\ref{#1})}

\def\derp#1#2{\rp{\partial{#1}}{\partial{#2}}}
\def\dert#1#2{\frac{{{\textrm{d}}}{#1}}{{{\textrm{d}}}{#2}}}

\def\virg#1{``#1"}

\def\eqi{\begin{equation}}
\def\eqf{\end{equation}}
\def\eqia{\begin{eqnarray}}
\def\eqfa{\end{eqnarray}}

\def\rp#1#2{\frac{#1}{#2}}
\def\lb#1{\label{#1}}

\def\bds#1{\boldsymbol{#1}}

\def\ton#1{\left(#1\right)}
\def\qua#1{\left[#1\right]}
\def\grf#1{\left\{#1\right\}}
\def\ang#1{\left\langle #1\right\rangle}
\documentclass[onecolumn]{aastex}

\usepackage{morefloats}
\usepackage[title]{appendix}
\usepackage{textcomp}
\usepackage{booktabs}
\usepackage[table,xcdraw]{xcolor}
\usepackage{multirow}
\usepackage{rotating,tabularx}
\usepackage{float}
\usepackage{enumerate}
\usepackage{rotating}
\usepackage[polutonikogreek,english]{babel}
\usepackage{amsmath,starfont,textgreek,w-greek,wasysym}
\usepackage[flushleft]{threeparttable}
\usepackage{amsthm}
\usepackage{amscd,lineno}
\usepackage{amssymb,dsfont}
\usepackage{graphicx,epsfig}
\usepackage{txfonts}
\usepackage{xr-hyper}
\usepackage{hyperref}

\RequirePackage{color}

\newcommand{\grk}[1]{\selectlanguage{polutonikogreek}
#1\selectlanguage{english}}

\allowdisplaybreaks[1]

\begin{document}

\title{One EURO for Uranus: the Elliptical Uranian Relativity Orbiter mission}

\shortauthors{L. Iorio, A.P. Girija, D. Durante}

\author{Lorenzo Iorio\altaffilmark{1}}
\affil{Ministero dell' Istruzione e del Merito
\\ Viale Unit\`{a} di Italia 68, I-70125, Bari (BA),
Italy}

\altaffiltext{1}{Corresponding author. Email: \texttt{lorenzo.iorio@libero.it}}

\author{Athul Pradeepkumar Girija}
\affil{Purdue University, West Lafayette, IN 47907, USA}

\author{Daniele Durante}
\affil{Department of Mechanical and Aerospace Engineering, Sapienza University of Rome, Rome, Italy}


\begin{abstract}
Recent years have seen increasing interest in sending a mission to Uranus, visited so far only by Voyager 2 in 1986.
EURO (Elliptical Uranian Relativity Orbiter) is a preliminary mission concept investigating the possibility of dynamically measuring the planet's angular momentum  by means of the Lense-Thirring effect affecting a putative Uranian orbiter. It is possible, at least in principle, to separate the relativistic precessions of the orbital inclination to the Celestial Equator and of the longitude of the ascending node  of the spacecraft from its classical rates of the pericentre induced by the multipoles of the planet's gravity field by adopting \textcolor{black}{an appropriate orbital configuration}. For a wide and elliptical $2\,000\times 100\,000\,\mathrm{km}$ orbit, the gravitomagnetic signatures amount to tens of milliarcseconds per year, while, for a suitable choice of the initial conditions, the peak-to-peak amplitude of the range-rate shift can reach the level of $\simeq 1.5\times 10^{-3}$ millimetre per second in a single pericentre passage of a few hours. By lowering the apocentre height to $10\,000\,\mathrm{km}$, the Lense-Thirring precessions are enhanced to the level of hundreds of milliarcseconds per year. The uncertainties in the orientation of \textcolor{black}{the planetary spin axis} and in \textcolor{black}{the inclination} are major sources of systematic bias; it turns out that they should be determined with accuracies as good as $\simeq 0.1-1$ and $\simeq 1-10$ milliarcseconds, respectively.
\end{abstract}


\section{Introduction}
So far, Uranus \citep{Uranus91,Hubbard1997}, the seventh planet of our solar system discovered in the eighteenth century \citep{1781RSPT...71..492H}, has been visited by an automated spacecraft only once when, in January 1986, the NASA probe Voyager 2 flew closely past it \citep{1986Sci...233...39S,1987JGR....9214873S}.

Recent years have seen a renewed interest for the exploration of Uranus and, possibly, Neptune\footnote{In general, missions to Uranus are favored due to logistical and cost reasons.} \citep{Neptune95}, discovered in the mid of the nineteenth century \citep{LeV46,1846MNRAS...7..153G}, boosting a number of investigations of possible spacecraft-based missions, submitted mostly to NASA and ESA, targeted to at least one or both the two icy giants \citep{rept11,2018P&SS..155...12M,rept18,2020P&SS..19105030F,2020Natur.579...17G,ESA21,rept22,2022Natur.604..607W,2019P&SS..17704680H,2021ExA...tmp..139G}. Their main targets are a better knowledge of the fundamental physical parameters of the most distant Sun's planets, such as their gravitational and magnetic fields, rotation rates, and deep atmospheric composition and temperature \citep{2020SSRv..216...38H}, and of their natural satellites, many of them may be icy ocean worlds that could possibly harbor life \citep{2017PNAS..114.4566M}.
At present, none of such proposals was approved by any space agency.

\textcolor{black}{Here, we make a cursory overview of some of the  past proposals. Among them}, there was the Uranus Pathfinder concept for an Uranian orbiter \citep{2012ExA....33..753A}, submitted in 2011 to ESA as a M-class mission. ODINUS (Origins, Dynamics, and Interiors of the Neptunian and Uranian Systems) was proposed to the ESA's Cosmic Vision programme in 2013 as a L-class mission aimed at sending two twin orbiters around Uranus and Neptune, respectively \citep{2014P&SS..104...93T}. The science case for an orbital mission to Uranus only, rooted in the ODINUS concept, was explored in a separated study \citep{2014P&SS..104..122A}. MUSE (Mission to Uranus for Science and Exploration) was another proposal to ESA for a L-class mission to Uranus \citep{2015AdSpR..55.2190B}. A further mission concept implying an Uranian orbiter, proposed to NASA in 2017, was OCEANUS (Origins and Composition of the Exoplanet Analog Uranus System) \citep{2017AdSpR..59.2407M}. A multi-probe mission to Uranus, including a SNAP (Small Next-Generation Atmospheric Probe), was studied by \citet{2020SSRv..216...72S}. QUEST (Quest to Uranus to Explore Solar System Theories) is a lower cost option for a flagship mission aimed to insert an orbiter around Uranus arisen during the 30th Annual NASA/JPL Planetary Science Summer Seminar \citep{2020AcAau.170....6J}.

At the time of writing, it seems that the latest proposal for a mission targeted at Uranus, submitted to NASA, is UOP (Uranus Orbiter and Probe) \citep{UOP21,UOP23}; it is giving rise to a number of related studies \citep{2022PSJ.....3...58C,2023AcAau.202..104G}. Preliminary hints of a possible flyby of Uranus with a future Chinese interplanetary exploration mission beyond Jupiter can be found in\footnote{See also https://www.space.com/china-probes-jupiter-uranus-same-launch on the Internet.} \citet{cinesi18}.

Among the physical parameters of interest of Uranus (and Neptune), whose determination/constraint is the main goal of the aforementioned proposed missions, there are the (normalized) moment of inertia (MoI) $\mathcal{J}$ and the rotation period $P$ \citep{2020RSPTA.37890474H,2020SSRv..216...38H,2022MNRAS.512.3124N}. In general, they depend on the distribution of matter in the interior of a planet, concurring to form its spin angular momentum
\eqi
S = \mathcal{J}\,M\,R^2\,\upomega,\lb{Spin}
\eqf
where $M$ is the planet's mass, $R$ is its equatorial radius, and
\eqi
\upomega\doteq \rp{2\uppi}{P}
\eqf is its rotational angular speed.
\citet{2022MNRAS.512.3124N} pointed out that a determination of the Uranian MoI $\mathcal{J}_{\uranus}$ at a $\simeq 1-0.1$ per cent level could constrain the planetary rotation period and the depth of the winds, respectively. \citet{2022MNRAS.512.3124N} acknowledged that it is not an easy task which could be implemented only with a future dedicated space mission.

General relativity\footnote{For a recent overview of the Einstein's theory of gravity, see, e.g., \citet{2016Univ....2...23D}, and references therein.} \citep{2017grav.book.....M} offers, in principle, a way to measure the spin angular momentum $\bds S$ of a rotating body in a dynamical, model-independent way. Indeed, in its weak-field and slow-motion approximation, its linearized equations formally resemble those of the Maxwellian electromagnetism giving rise to the so-called \virg{gravitoelectromagnetic} paradigm. For such a concept within the framework of the Einsteinian theory of gravitation, see, e.g., \citet{1958NCim...10..318C,Thorne86,1986hmac.book..103T,1988nznf.conf..573T,1991AmJPh..59..421H,
1992AnPhy.215....1J,2001rfg..conf..121M,2001rsgc.book.....R,Mash07,2008PhRvD..78b4021C,
2014GReGr..46.1792C,2021Univ....7..388C,2021Univ....7..451R}, and references therein.
Actually, general relativistic gravitoelectromagnetism has nothing to do with electric charges and currents, implying a number of purely gravitational phenomena affecting orbiting test particles, precessing gyroscopes, moving clocks and atoms, and propagating electromagnetic waves \citep{1977PhRvD..15.2047B,1986SvPhU..29..215D,2002EL.....60..167T,2002NCimB.117..743R,2004GReGr..36.2223S,2009SSRv..148...37S,2009SSRv..148..105S}.
In particular, it turns out that matter-energy currents give rise to a \virg{gravitomagnetic} component of the gravitational field encoded in the off-diagonal components $g_{0i},\,i=1,\,2,\,3$ of the spacetime metric tensor $g_{\upmu\upnu},\,\upmu,\,\upnu=0,\,1,\,2,\,3$. To the first post-Newtonian (1pN) order, in the case of an isolated, slowly spinning body, the source of its gravitomagnetic field is just its spin angular momentum $\bds S$ which, among other things, induces a non-central, Lorentz-like acceleration on an orbiting test particle. It causes secular precessions of the orbit of the latter  \citep{Sof89,1991ercm.book.....B,SoffelHan19} which go by the name of Lense-Thirring (LT) effect \citep{1918PhyZ...19..156L,1984GReGr..16..711M}. Gravitomagnetism has been experimentally measured in a, so far, undisputed way only in the field of the spinning Earth with the dedicated GP-B spacecraft-based mission which measured the Pugh-Schiff precessions \citep{Pugh59,Schiff60} of the axes of four gyroscopes carried onboard to a 19 per cent accuracy  \citep{2011PhRvL.106v1101E}. As per the LT orbital precessions, somewhat controversial attempts to measure them with the Earth's artificial satellites of the LAGEOS type \citep{2019JGeod..93.2181P} and the Satellite Laser Ranging (SLR) technique \citep{SLR11} are currently ongoing \citep{2013NuPhS.243..180C,2013CEJPh..11..531R,2013AcAau..91..141I}; see \citet{2011Ap&SS.331..351I} and references therein also for other proposed tests with natural and artificial bodies in the solar system. Recently, a successful detection of the gravitomagnetic orbital precessions in a tight astrophysical  binary system made of a white dwarf and a pulsar was claimed \citep{2020Sci...367..577V}, but also such a test subsequently raised concerns \citep{2020MNRAS.495.2777I}. Efforts for measuring the LT periastron precession of the double pulsar PSR J0737-3039A/B \citep{2003Natur.426..531B,2004Sci...303.1153L} in the next future are underway \citep{Kehletal017,2020MNRAS.497.3118H}.

\textcolor{black}{Here, we investigate the possibility} of measuring the gravitomagnetic LT orbital precessions of a putative Uranian orbiter, provisionally dubbed EURO (Elliptical Uranian Relativity Orbiter), independently of the competing ones due to the even and odd zonal harmonics $J_\ell,\,\ell=2,\,3,\,4,\,\ldots$ of the planet's gravity field, which are other physical parameters of great interest \citep{2022MNRAS.512.3124N}. \textcolor{black}{The peculiar obliquity of $98^\circ$ of the Uranian spin axis \citep{Uranus91,Hubbard1997}, whose origin is, at present,  actively investigated \citep{2021PSJ.....2...78R,2022A&A...668A.108S}, is instrumental for the achievement of such a goal.}
For the sake of clarity, the standard Keplerian orbital elements \citep{2000ssd..book.....M,Bertotti03,2011rcms.book.....K} are used to perform a sensitivity analysis\textcolor{black}{,} which however is not intended to replace future simulations of the actual observables \textcolor{black}{(such as range-rate), including} fitting parameters and a full covariance analysis.

The paper is organized as follows.
In Section\,\ref{sec2}, the LT effect for an arbitrary orientation of the primary's spin axis is reviewed, while the competing classical precessions induced by the quadrupole mass moment of the central body are treated in  Section\,\ref{sec3}. Section\,\ref{sec4} deals with the orbital geometry which better allows to separate the aforementioned relativistic and Newtonian effects each other. In Section\,\ref{sec5}, numerical values are given for a pair of specific wide and elliptical trajectories of EURO, and the range-rate is considered as well. The impact of the errors in the planetary spin axis orientation and in the orbital inclination on the classical precessions is examined. In Section\,\ref{sec6}, the conditions and the requirements for the implementation of the previously considered orbits are discussed. Section\,\ref{sec7} summarizes our findings and offers our conclusions. Appendix\,\ref{appenA} shows the details of the calculation of the classical orbital precessions due to the first seven zonal harmonics of the planet's gravity field, while  Appendix\,\ref{appenB} is dedicated to the approximate analytical calculation of the range-rate shift of a spacecraft orbiting a distant planet. In Appendix\,\ref{appenC}, some suggestions for properly naming the pericentre and the apocentre in the case of Uranus are offered.
\section{The generalized Lense-Thirring orbital precessions}\lb{sec2}
Let an isolated, massive rotating body orbited by a test particle be considered.

The generalized\footnote{\lb{nota1}The formulas obtained originally by \citet{1918PhyZ...19..156L} and often reported in the literature \citep{Sof89,1991ercm.book.....B,SoffelHan19} hold when the orientation of $\bds S$ is known, so that it can be aligned with, say, the reference $z$ axis of the coordinate system adopted; in such a case, the reference $\grf{x,\,y}$ plane coincides with the body's equatorial one.} LT averaged rates of the inclination $I$ of the orbital plane to the reference $\grf{x,\,y}$ plane adopted, of the longitude of the ascending node $\Omega$ and of the argument of pericentre $\omega$ are \citep{2017EPJC...77..439I}
\begin{align}
\dot I^\mathrm{LT} \lb{ILT} & = \rp{2\,G\,S\,\ton{\bds{\hat{k}}\bds\cdot\bds{\hat{l}}}}{c^2\,a^3\ton{1-e^2}^{3/2}}, \\ \nonumber \\
\dot\Omega^\mathrm{LT} \lb{OLT} & = \rp{2\,G\,S\,\csc I\,\ton{\bds{\hat{k}}\bds\cdot\bds{\hat{m}}}}{c^2\,a^3\ton{1-e^2}^{3/2}}, \\ \nonumber \\
\dot\omega^\mathrm{LT} \lb{oLT} & = -\rp{2\,G\,S\,\bds{\hat{k}}\bds\cdot\ton{2\,\bds{\hat{h}} + \cot I\,\bds{\hat{m}}}}{c^2\,a^3\ton{1-e^2}^{3/2}};
\end{align}
for other derivations based on different parameterizations, see also \citet{1975PhRvD..12..329B,1988NCimB.101..127D,1992PhRvD..45.1840D}.
In \rfrs{ILT}{oLT},  $G$ is the Newtonian constant of gravitation, $c$ is the speed of light in vacuum,  $\bds{\hat{k}}$ is the spin axis of the central body so that $\bds S = S\,\bds{\hat{k}}$,
\eqi
\bds{\hat{h}}=\grf{\sin I\,\sin\Omega,\,-\sin I\,\cos\Omega,\,\cos I}\lb{hvec}
\eqf is the unit vector of the orbital angular momentum of the test particle \citep{Sof89,1991ercm.book.....B,SoffelHan19},
\eqi
\bds{\hat{l}}=\grf{\cos\Omega,\,\sin\Omega,\,0}\lb{lvec}
\eqf
is the unit vector directed along the line of the nodes toward the ascending node \citep{Sof89,1991ercm.book.....B,SoffelHan19},
\eqi
\bds{\hat{m}}=\grf{-\cos I\,\sin\Omega,\,\cos I\,\cos\Omega,\,\sin I}\lb{mvec}
\eqf
is a unit vector lying in the orbital plane \citep{Sof89,1991ercm.book.....B,SoffelHan19} such that $\bds{\hat{l}}\bds\times\bds{\hat{m}} = \bds{\hat{h}}$,  $a$ is the semimajor axis, and $e$ is the eccentricity.

Since the reference frame usually adopted for processing solar system's spaceraft data is the ICRF (International Celestial Reference Frame) \citep{1998AJ....116..516M,iers10}, whose reference plane, the Celestial Equator (CE), coincides with the Earth's equatorial plane at the reference epoch J2000, it is useful to parameterize $\bds{\hat{k}}$ in terms of the right ascension (RA) $\alpha$ and declination (DEC) $\delta$ of the planet's pole of rotation as
\begin{align}
{\hat{k}}_x \lb{Kx} & = \cos\delta\cos\alpha, \\ \nonumber \\
{\hat{k}}_y \lb{Ky} & = \cos\delta\sin\alpha, \\ \nonumber \\
{\hat{k}}_z \lb{Kz} & = \sin\delta.
\end{align}
By means of \rfrs{Kx}{Kz}, \rfrs{ILT}{oLT} can be cast into the form
\begin{align}
\dot I^\mathrm{LT} \lb{iLT}& = \rp{2\,G\,S\,\cos\delta\cos\eta}{c^2\,a^3\,\ton{1-e^2}^{3/2}}, \\ \nonumber \\
\dot \Omega^\mathrm{LT} \lb{nLT}& = \rp{2\,G\,S\,\ton{\sin\delta + \cos\delta\,\cot I\,\sin\eta}}{c^2\,a^3\,\ton{1-e^2}^{3/2}}, \\ \nonumber \\
\dot \omega^\mathrm{LT} \lb{pLT}& = -\rp{2\,G\,S\,\qua{3\cos I\,\sin\delta + \cos\delta\,\ton{\csc I - 3\sin I}\,\sin\eta}}{c^2\,a^3\,\ton{1-e^2}^{3/2}},
\end{align}
where
\eqi
\eta\doteq \alpha-\Omega
\eqf
is the planetary spin's RA reckoned from the line of the nodes. Note that, for
\eqi
\delta = 90^\circ,
\eqf
the usual formulas by \citet{1918PhyZ...19..156L},
\begin{align}
\dot I^\mathrm{LT} \lb{kiz1}&=0, \\ \nonumber \\
\dot\Omega^\mathrm{LT} & = \rp{2\,G\,S}{c^2\,a^3\,\ton{1-e^2}^{3/2}}, \\ \nonumber\\
\dot\omega^\mathrm{LT} \lb{kiz3}& = -\rp{6\,G\,S\,\cos I}{c^2\,a^3\,\ton{1-e^2}^{3/2}},
\end{align}
valid, e.g., for an Earth's satellite, are restored.

In the case of Uranus, the general formulas of \rfrs{ILT}{oLT}, or \rfrs{iLT}{pLT},  are required since the orientation of its spin axis with respect to the ICRF is given by \citep{2014AJ....148...76J}
\begin{align}
\alpha_{\uranus} \lb{RA}& = 77.310^\circ, \\ \nonumber \\
\delta_{\uranus} \lb{DEC}& = 15.172^\circ.
\end{align}
Thus,
\begin{align}
{\hat{k}}^{\uranus}_x \lb{kx} & =  0.212, \\ \nonumber \\
{\hat{k}}^{\uranus}_y \lb{ky} & = 0.941, \\ \nonumber \\
{\hat{k}}^{\uranus}_z \lb{kz} & = 0.262;
\end{align}
the spin of Uranus mainly lies in the CE.
As a consequences, \rfrs{kiz1}{kiz3} cannot be used in the present case.
\section{The orbital precessions induced by $J_2$}\lb{sec3}
The first even zonal harmonic coefficient $J_2$ of the multipolar expansion of the classical part of the non-spherical planetary gravitational potential $U\ton{\bds r}$ induces the largest orbital precessions among those due to the deviations of $U$ from spherical symmetry \citep{Capde05}; they act as major systematic bias with respect to the relativistic rates of \rfrs{ILT}{oLT}.

For an arbitrary\footnote{The same considerations as in footnote\,\ref{nota1} hold here for the oblateness-induced orbital precessions. Their expressions for $\bds{\hat{k}}$ aligned with the reference $z$ axis can be found, e.g., in \citet{Capde05}; they are reported in \rfrs{ozi}{eza}.} orientation of $\bds{\hat{k}}$ in space, the oblateness-driven averaged rates are \citep{2017EPJC...77..439I}
\begin{align}
\dot I^{J_2} \lb{IJ2} & = -\rp{3}{2}\,\nk\,J_2\,\ton{\rp{R}{p}}^2\,\ton{\bds{\hat{k}}\bds\cdot\bds{\hat{l}}}\,\ton{\bds{\hat{k}}\bds\cdot\bds{\hat{h}}}, \\ \nonumber \\
\dot\Omega^{J_2} \lb{OJ2} & = -\rp{3}{2}\,\nk\,J_2\,\ton{\rp{R}{p}}^2\,\csc I\,\ton{\bds{\hat{k}}\bds\cdot\bds{\hat{m}}}\,\ton{\bds{\hat{k}}\bds\cdot\bds{\hat{h}}}, \\ \nonumber \\
\dot\omega^{J_2} \lb{oJ2} & = \rp{3}{4}\,\nk\,J_2\,\ton{\rp{R}{p}}^2\,\grf{2 -3\,\qua{\ton{\bds{\hat{k}}\bds\cdot\bds{\hat{l}}}^2 + \ton{\bds{\hat{k}}\bds\cdot\bds{\hat{m}}}^2} + 2\,\cot I\,\ton{\bds{\hat{k}}\bds\cdot\bds{\hat{m}}}\,\ton{\bds{\hat{k}}\bds\cdot\bds{\hat{h}}}},
\end{align}
where
\eqi
\nk\doteq \sqrt{\rp{GM}{a^3}}
\eqf
is the Keplerian mean motion, and
\eqi
p\doteq a\,\ton{1-e^2}
\eqf
is the orbital semilatus rectum.

By means of \rfrs{Kx}{Kz}, \rfrs{IJ2}{oJ2} can be suitably expressed as
\begin{align}
\dot I^{J_2}\lb{iobl} & = \rp{3}{2}\,\nk\,J_2\,\ton{\rp{R}{p}}^2\,\qua{\cos\delta\cos\eta\,\ton{-\cos I\,\sin\delta + \cos\delta\,\sin I\,\sin\eta}},\\ \nonumber\\
\dot\Omega^{J_2} \lb{Oobl}& = -\rp{3}{2}\,\nk\,J_2\,\ton{\rp{R}{p}}^2\,\csc I\,\ton{\sin I\,\sin\delta + \cos I\,\cos\delta\,\sin\eta}\,\ton{\cos I\,\sin\delta -
   \cos\delta\,\sin I\,\sin\eta},\\ \nonumber \\
\dot\omega^{J_2} \nonumber \lb{oobl}& = -\rp{3}{8}\,\nk\,J_2\,\ton{\rp{R}{p}}^2\,\qua{-4 + \ton{1 - 5\,\cos 2I}\,\sin^2\delta + \ton{3 - 5\,\cos 2I}\,\cot I\,
\sin 2 \delta\,\sin\eta + \right.\\ \nonumber \\
&\left. + 2\,\cos^2\delta\,\ton{3\,\cos^2\eta + 5\,\cos^2 I\,\sin^2\eta}}.
\end{align}
Note that, for
\eqi
\delta = 90^\circ,
\eqf
\rfrs{iobl}{oobl} reduce to the standard formulas \citep{Capde05}
\begin{align}
\dot I^{J_2} \lb{ozi} & = 0, \\ \nonumber \\
\dot\Omega^{J_2} & = -\rp{3}{2}\,\nk\,J_2\,\ton{\rp{R}{p}}^2\,\cos I, \\ \nonumber \\
\dot\omega^{J_2} \lb{eza}& = \rp{3}{8}\,\nk\,J_2\,\ton{\rp{R}{p}}^2\,\ton{3 + 5\cos 2 I}.
\end{align}
valid, e.g., for an Earth's satellite.

The details of the calculation of the precessions induced by the zonal harmonics $J_\ell$ of any degree $\ell$ can be found in Appendix\,\ref{appenA}.
\section{Choosing the most suitable orbital geometry}\lb{sec4}
From \rfrs{ILT}{oLT} and \rfrs{IJ2}{oJ2}, it turns out that, if
\begin{align}
\bds{\hat{k}}\bds\cdot\bds{\hat{h}} \lb{kh} & =0, \\ \nonumber \\
\bds{\hat{k}}\bds\cdot\bds{\hat{m}} \lb{km} & \neq 0, \\ \nonumber \\
\bds{\hat{k}}\bds\cdot\bds{\hat{l}} \lb{kl} & \neq 0,
\end{align}
i.e., if the body's spin axis lies in the orbital plane \textcolor{black}{of the test particle} somewhere between the line of the nodes and the direction perpendicular to the latter,
the $J_2$-induced shifts of $I$ (\rfr{IJ2}) and $\Omega$ (\rfr{OJ2}) vanish, contrary to the corresponding gravitomagnetic ones of \rfrs{ILT}{OLT}. Then, the relativistic node and inclination precessions would not be biased at all by the classical disturbances due to the body's oblateness. The condition of \rfr{kh}, corresponding to a polar orbit, is usually selected in order to fulfil some of the planetological scopes of most of the proposed Uranian missions (see, e.g., \citet{UOP21,2023AcAau.202..104G}).

By means of  \rfrs{hvec}{Kz}, the conditions of \rfrs{kh}{kl} are equivalent to
\begin{align}
\tan I \lb{cond1} & = \csc\eta\,\tan\delta, \\ \nonumber \\
\tan I \lb{cond2} &\neq -\sin\eta\,\cot\delta, \\ \nonumber\\
\cos\eta\,\cos\delta \lb{cond3} &\neq 0.
\end{align}
In the case of Uranus, according to \rfr{DEC}, it is
\begin{align}
\sin\delta_{\uranus} & = 0.262\neq 0, \\ \nonumber \\
\cos\delta_{\uranus} & = 0.965\neq 0.
\end{align}
Thus, for EURO, \rfrs{cond1}{cond3} can be simultaneously fulfilled if \textcolor{black}{the planetary spin's RA relative to the node}
\eqi
\eta\neq 90^\circ.
\eqf

In particular, if one chooses
\eqi
\eta = 0,\lb{et}
\eqf
then, from \rfrs{cond1}{cond3}, it turns out that
\eqi
I = 90^\circ,\lb{inkl}
\eqf
implying that the gravitomagnetic pericentre precession of \rfr{pLT} vanishes, contrary to the classical one of \rfr{oobl}.
Indeed, \rfrs{iLT}{nLT} and \rfr{oobl}, calculated with \rfrs{et}{inkl}, reduce to the following secular rates
\begin{align}
\dot I^\mathrm{LT} \lb{iLT2}& = \rp{2\,G\,S\,\cos\delta}{c^2\,a^3\,\ton{1-e^2}^{3/2}}, \\ \nonumber \\
\dot \Omega^\mathrm{LT} \lb{nLT2}& = \rp{2\,G\,S\,\sin\delta}{c^2\,a^3\,\ton{1-e^2}^{3/2}}, \\ \nonumber \\
\dot\omega^{J_2} \lb{oobl2}& = -\rp{3}{4}\,\nk\,J_2\,\ton{\rp{R}{p}}^2.
\end{align}
Furthermore, as shown in Appendix\,\ref{appenA}, the averaged classical rates of $I$ and $\Omega$ vanish also for the even and odd zonals of degree $\ell> 2$, contrary to the pericentre which turns out to be impacted  by the odd zonals as well.
Thus, in principle, one could separate the general relativistic signatures of $I$ and $\Omega$ from the Newtonian ones of $\omega$ due to $J_\ell,\,\ell=2,\,3,\,4,\ldots$ which could be used to simultaneously determine the multipolar field of Uranus without any a-priori \virg{imprinting} due to its 1pN gravitomagnetic field\textcolor{black}{, and vice-versa: $\bds{S}$ and $J_\ell,\,\ell=2,\,3,\,4,\ldots$ could, then, be measured independently of each other. It would be important since, although there is \textcolor{black}{no strict} one-to-one correspondence between $J^{\uranus}_2$ and the Uranus' MoI \citep[Sect.\,3.3]{2022MNRAS.512.3124N}, there is a rather strong correlation between the other Uranian multipolar coefficients $J_4,\,J_6,\,J_8$, and the MoI \citep{2022MNRAS.512.3124N}.} General relativity would, instead, imprint the classical pericentre precessions through the mismodelled part of its \virg{gravitoelectric} 1pN counterpart given by
\citep{Sof89,1991ercm.book.....B,SoffelHan19}
\eqi
\dot\omega^\mathrm{GE} = \ton{\rp{2+2\,\gamma-\beta}{3}}\rp{3\,\nk\,\mu}{c^2\,p},\lb{oGE}
\eqf
where $\gamma,\,\beta$ are the parameterized post-Newtonian (PPN) parameters \citep{1972ApJ...177..757W,2018tegp.book.....W} which are equal to unity in general relativity.
\section{The case of a polar, highly eccentric Uranian orbiter}\lb{sec5}
By \textcolor{black}{calculating \rfr{Spin} with \citep[Tab.\,1]{2022MNRAS.512.3124N}}
\eqi
\textcolor{black}{P_{\uranus} = 17.24\,\mathrm{hr}}
\eqf
\textcolor{black}{and \citep[Tab.\,2]{2022MNRAS.512.3124N}}
\eqi
\textcolor{black}{\mathcal{J}_{\uranus} = 0.22594,}
\eqf
along with the reference radius for the Uranian zonal harmonics \citep{1988Icar...73..349F}
\eqi
R_{\uranus}=25\,559\,\mathrm{km}\lb{raggio}
\eqf
\textcolor{black}{and\textcolor{black}{\footnote{\textcolor{black}{The values of \rfrs{raggio}{muura} are reported also in  \citet[Tab.\,1]{2022MNRAS.512.3124N} }}} \citep{2014AJ....148...76J}}
\eqi
\textcolor{black}{\mu_{\uranus} = 5\,794\,556.4\times 10^9\,\mathrm{m}^3\,\mathrm{s}^{-2},}\lb{muura}
\eqf
a reasonable guess for the Uranian spin angular momentum may be
\eqi
S_{\uranus}\simeq 1.29\times 10^{36}\,\mathrm{kg}\,\mathrm{m}^2\,\mathrm{s}^{-1}.\lb{Sura}
\eqf
\textcolor{black}{It turns out that calculating \rfr{Spin} with the other values for $P_{\uranus}$ and $\mathcal{J}_{\uranus}$ in \citet[Tab.\,1-2]{2022MNRAS.512.3124N} yields a $\simeq 1$ per cent uncertainty in the theoretically predicted value of $S_{\uranus}$.
}

As far as the shape and the size of the spacecraft's path are concerned, wide and highly elliptical orbits are usually adopted for the orbital phase of several proposed missions. \textcolor{black}{Just a}s an example for Uranus, a $4\,000\times 550\,000\,\mathrm{km}$ orbit was recently proposed by \citet{2023AcAau.202..104G} in several simulations.

\begin{figure}[h]
\centering
\begin{tabular}{c}
\includegraphics[width = 16 cm]{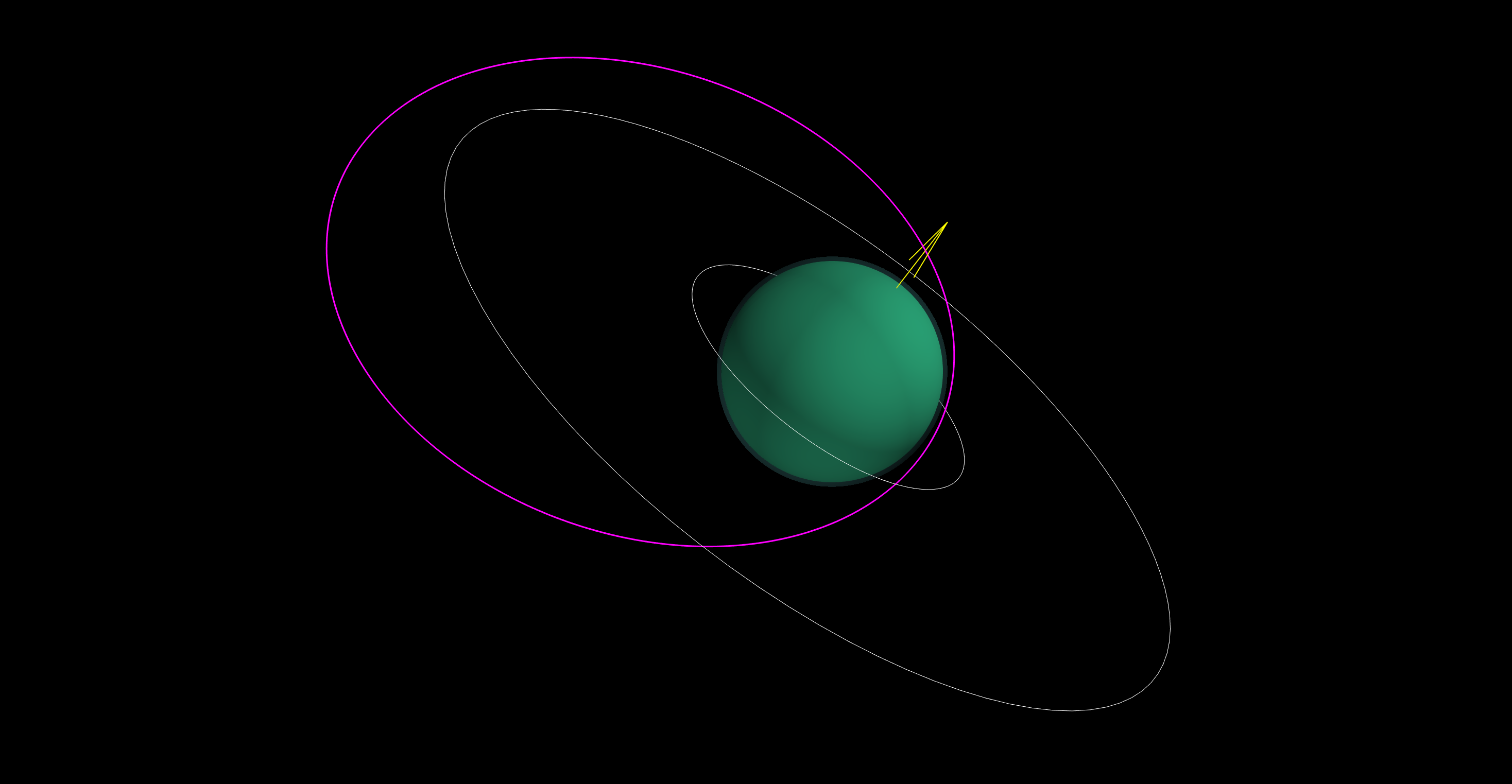}\\
\end{tabular}
\caption{
Orbital geometry of EURO for a $2\,000\times 100\,000\,\mathrm{km}$ orbit \textcolor{black}{(purple)} oriented with respect to the ICRF according to \rfrs{et}{inkl}. The Uranian spin axis, in yellow, is oriented as per \rfrs{RA}{DEC} with respect to the CE. \textcolor{black}{The white circles represent the innermost ($1986\mathrm{U}2\mathrm{R}/\upzeta$) and outermost ($\upmu$) rings of Uranus, respectively.}
}\label{fig0}
\end{figure}

Here, a lower and smaller $2\,000\times 100\,000\,\mathrm{km}$ path,  corresponding to
\eqi
a  = 76\,559\,\mathrm{km},\,e = 0.6400,\,\Pb = 15.3\,\mathrm{hr},\lb{per1}
\eqf
where $\Pb$ is the orbital period \textcolor{black}{of the probe}, will be examined. Figure\,\ref{fig0} depicts such kind of orbit \textcolor{black}{(purple)} oriented according to \rfrs{et}{inkl}.

\begin{figure}[h]
\centering
\begin{tabular}{c}
\includegraphics[width = 16 cm]{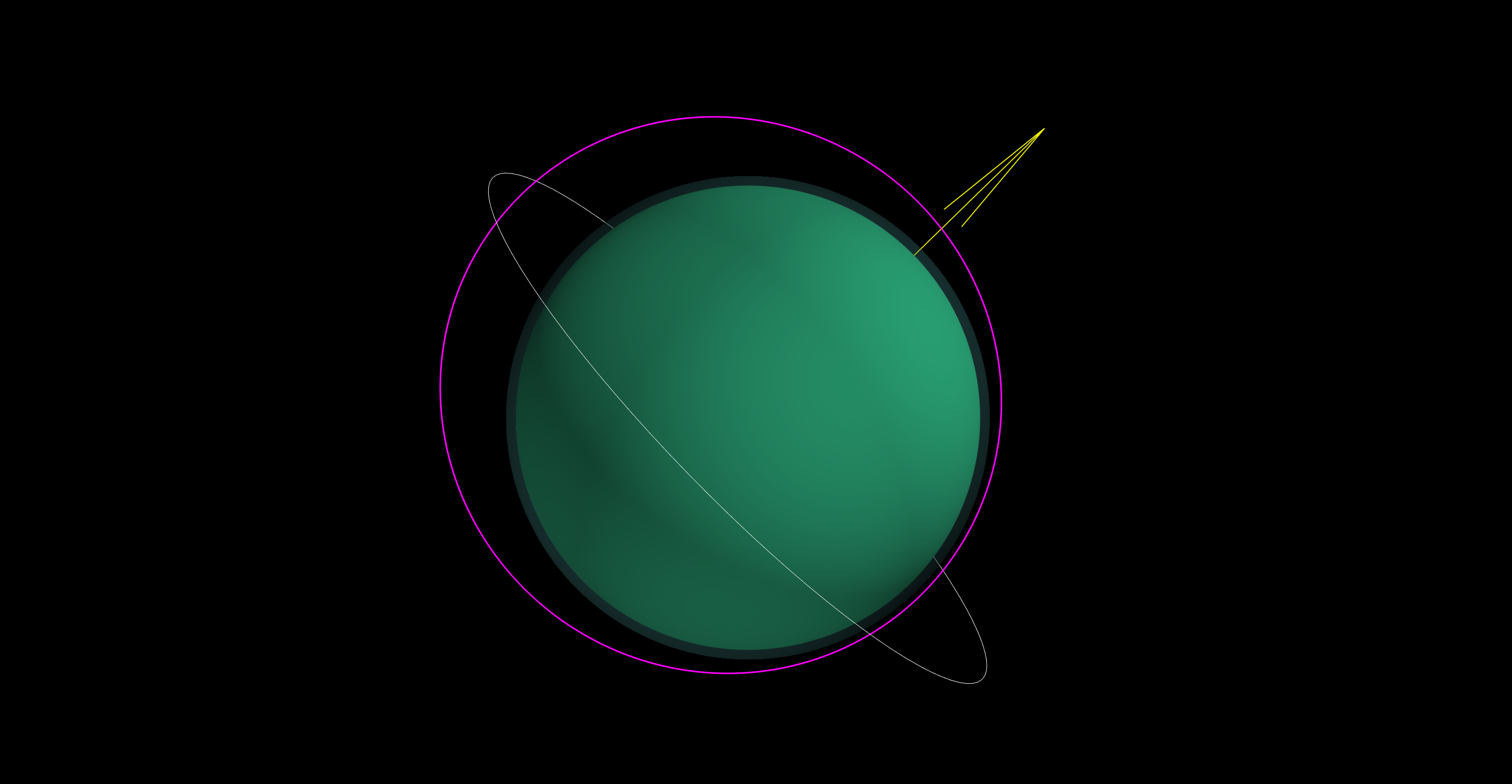}\\
\end{tabular}
\caption{
Orbital geometry of EURO for a $2\,000\times 10\,000\,\mathrm{km}$ orbit oriented with respect to the ICRF according to \rfrs{et}{inkl}. The Uranian spin axis, in yellow, is oriented as per \rfrs{RA}{DEC} with respect to the CE. \textcolor{black}{The white circles represent the innermost ($1986\mathrm{U}2\mathrm{R}/\upzeta$) and outermost ($\upmu$) rings of Uranus, respectively.}
}\label{fig1}
\end{figure}

A further scenario, shown in Figue\,\ref{fig1}, encompasses a less eccentric orbit with  apocentre height $h_\mathrm{apo} = 10\,000\,\mathrm{km}$, corresponding to
\eqi
\textcolor{black}{a = 31\,559\,\mathrm{km},\,e  = 0.1267,\,\Pb \lb{per2} = 4.06\,\mathrm{hr}.}
\eqf

For details on the practical implementation of such orbital geometries, see Section\,\ref{sec6}.
\subsection{The orbital precessions}\lb{orbpr}
By assuming \citep{2014AJ....148...76J}
\eqi
J_2^{\uranus}\ton{\times 10^6} = 3\,510.7,\lb{J2}
\eqf
\rfrs{iLT2}{oobl2} and\footnote{The gravitoelectric pericentre precession was computed for $\gamma = \beta = 1$.} \rfr{oGE}, calculated with the orbital parameters of \textcolor{black}{\rfr{per1}}, yield
\begin{align}
\dot I^\mathrm{LT} \lb{numILT} & \simeq 59.5\,\mathrm{mas\,yr}^{-1}, \\ \nonumber \\
\dot\Omega^\mathrm{LT} \lb{numOLT}& \simeq 16.1\,\mathrm{mas\,yr}^{-1}, \\ \nonumber \\
\dot\omega^{J_2} \lb{numoJ2}& = -173.0^\circ\,\mathrm{yr}^{-1}, \\ \nonumber \\
\dot\omega^\mathrm{GE} \lb{numoGE}& = 3.2\,\mathrm{arcsec\,yr}^{-1},
\end{align}
where $\mathrm{mas\,yr}^{-1}$ and $\mathrm{arcsec\,yr}^{-1}$ stand for milliarcseconds per year and arcseconds per year, respectively.

The $2\,000\times 10\,000\,\mathrm{km}$  orbit of \textcolor{black}{\rfr{per2}} yields
\begin{align}
\dot I^\mathrm{LT} \lb{kazILT} & \simeq 394.6\,\mathrm{mas\,yr}^{-1}, \\ \nonumber \\
\dot\Omega^\mathrm{LT} \lb{kazOLT}& \simeq 107.0\,\mathrm{mas\,yr}^{-1}, \\ \nonumber \\
\dot\omega^{J_2} \lb{kazoJ2}& = -1\,384.9^\circ\,\mathrm{yr}^{-1}, \\ \nonumber \\
\dot\omega^\mathrm{GE} \lb{kazoGE}& = 17.4\,\mathrm{arcsec\,yr}^{-1},
\end{align}

\begin{figure}[h]
\centering
\begin{tabular}{cc}
\includegraphics[width = 7.5 cm]{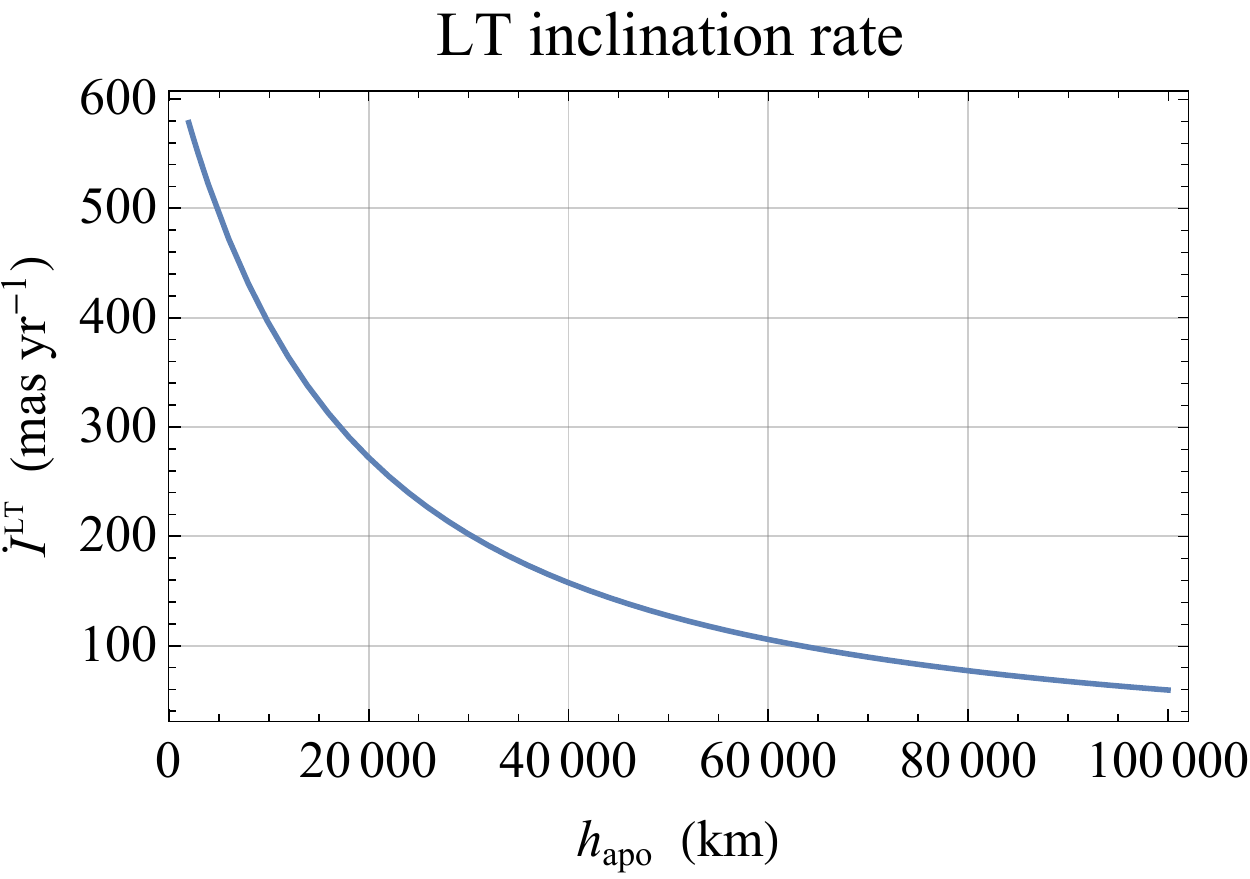} & \includegraphics[width = 7.5 cm]{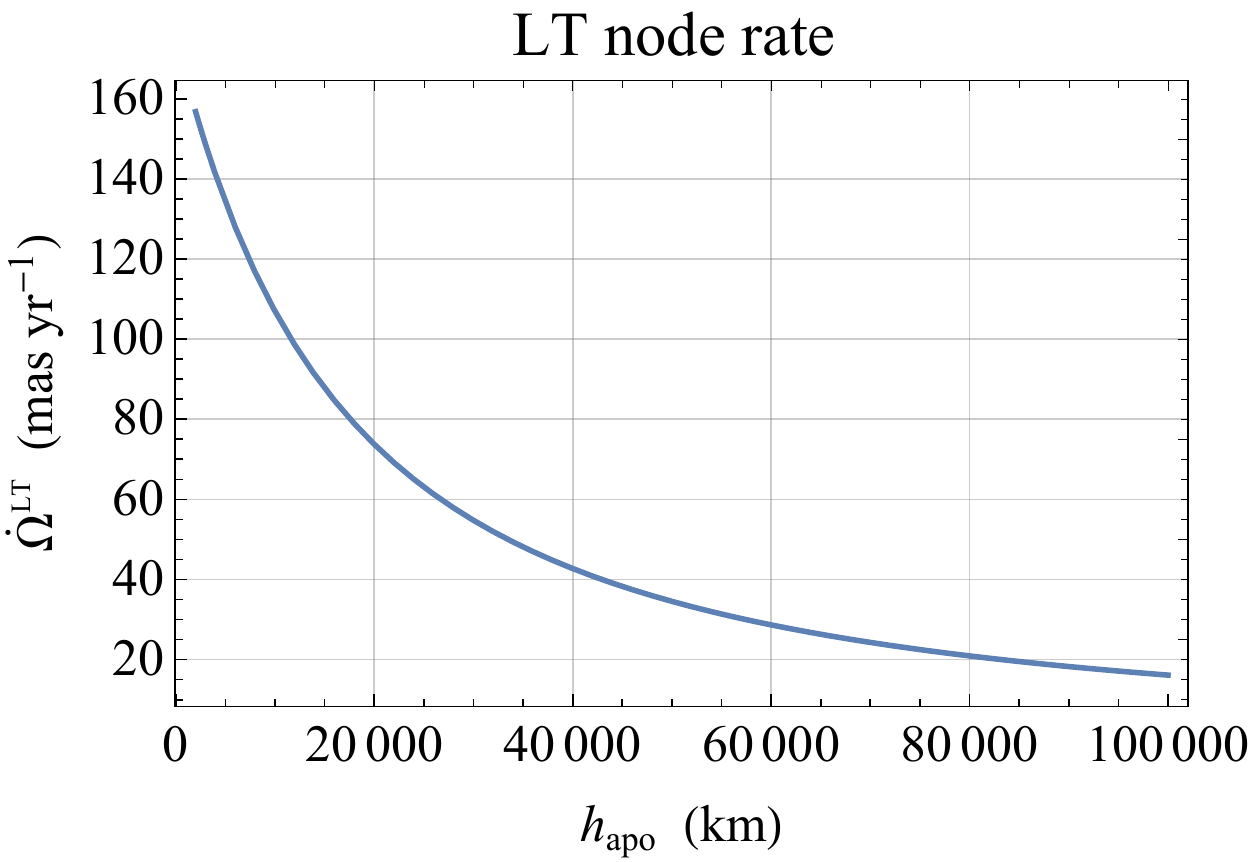}\\
\includegraphics[width = 7.5 cm]{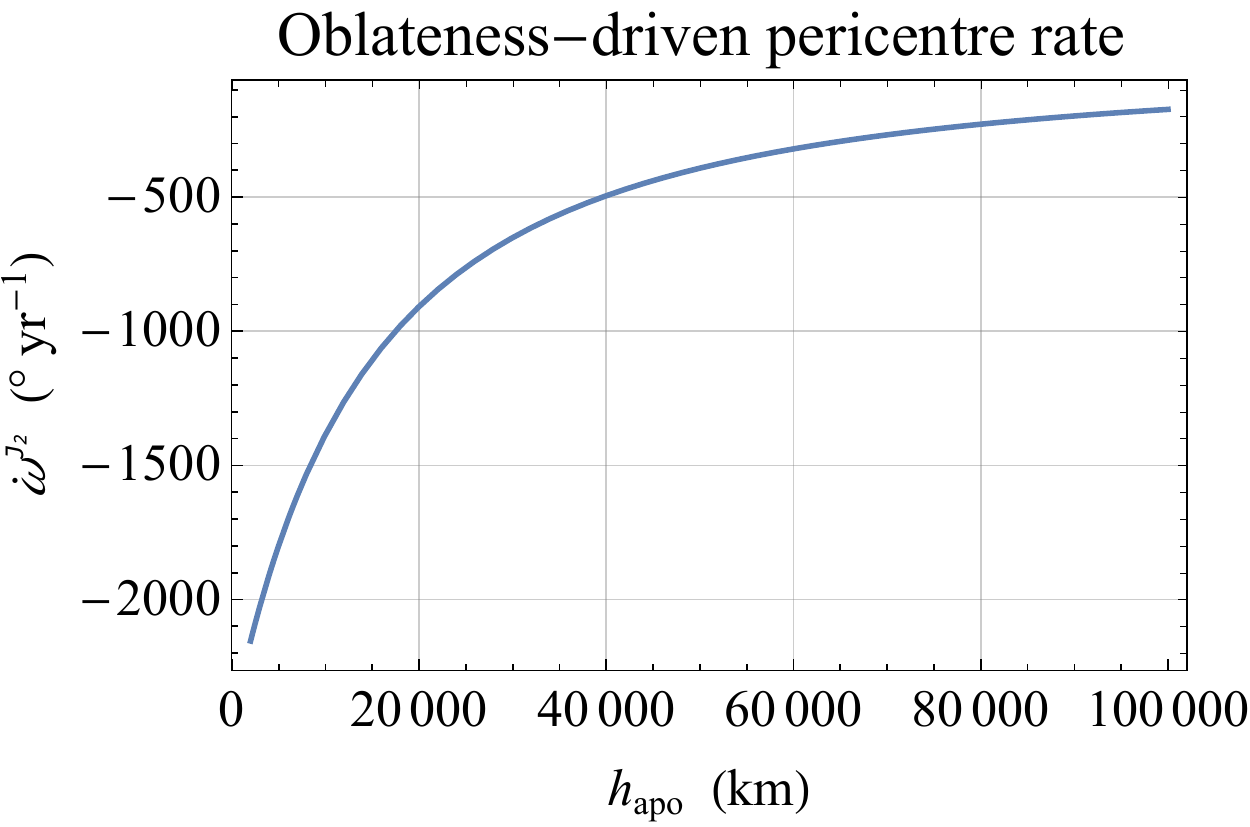} & \includegraphics[width = 7.5 cm]{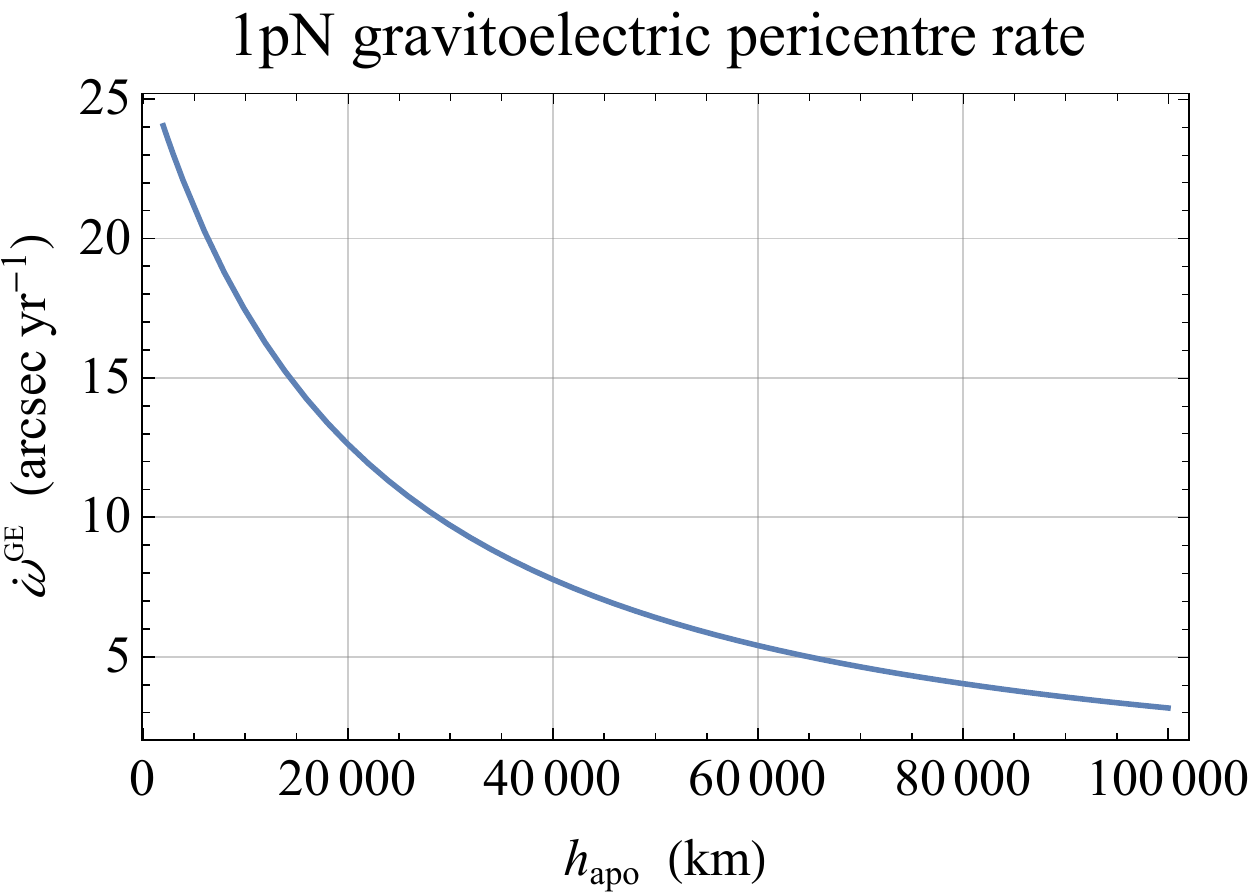}\\
\includegraphics[width = 7.5 cm]{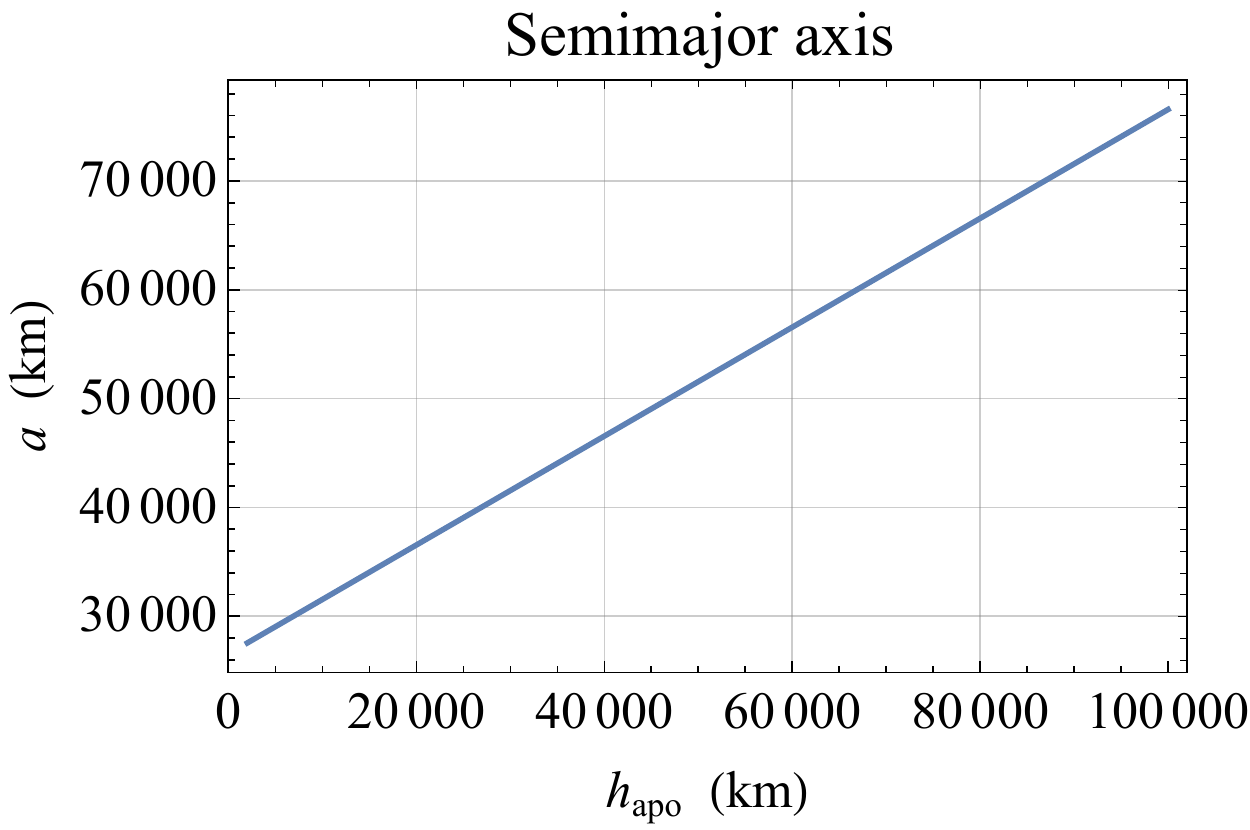} & \includegraphics[width = 7.5 cm]{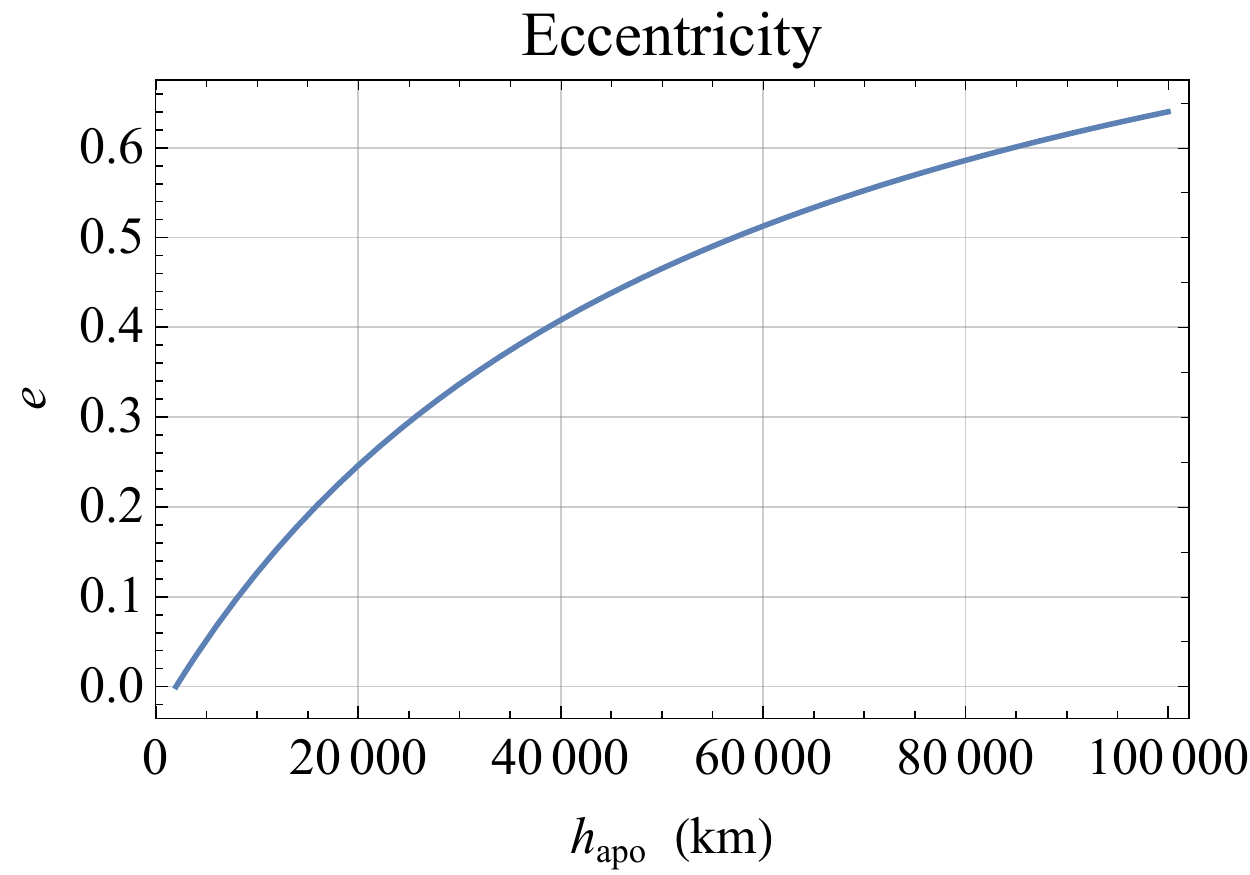}\\
\end{tabular}
\caption{
Upper row: LT precessions of the inclination $I$ (left panel) and of the node $\Omega$ (right panel), in mas yr$^{-1}$, calculated from \rfrs{iLT2}{nLT2} as functions of  $h_\mathrm{apo}$, in km, for a fixed value of  $h_\mathrm{peri}=2\,000\,\mathrm{km}$. For $\delta_{\uranus}$ and $S_{\uranus}$, \rfr{DEC} and \rfr{Sura} were adopted, respectively. Middle row: oblateness-driven (left panel) and 1pN gravitoelectric (right panel) precessions of the pericentre $\omega$, in degrees per year ($^\circ$ yr$^{-1}$) and arcseconds per year (arcsec yr$^{-1}$), respectively, calculated from \rfrs{oobl2}{oGE} as functions of  $h_\mathrm{apo}$, in km, for a fixed value of  $h_\mathrm{peri}=2\,000\,\mathrm{km}$. For $R_{\uranus}$ and $J_2^{\uranus}$, \rfr{raggio} and \rfr{J2} were adopted, respectively. Lower row: semimajor axis $a$, in km, (left panel) and eccentricity $e$ (right panel) calculated for as functions of  $h_\mathrm{apo}$, in km, for a fixed value of  $h_\mathrm{peri}=2\,000\,\mathrm{km}$.
}\label{fig2}
\end{figure}

Figure\,\ref{fig2} displays the nominal orbital precessions as per \rfrs{iLT2}{oGE} along with the semimajor axis $a$, in km, and the eccentricity $e$ as functions of $h_\mathrm{apo}$, ranging from 2\,000 to $100\,000$ km, for a fixed value of the pericentre height $h_\mathrm{peri}=2\,000\,\mathrm{km}$.
\clearpage
%
%
%

About the \virg{imprinting} of the mismodelling in the 1pN gravitolectric pericentre precession on the classical one, from \rfr{oGE} it turns out that, in principle, it is mainly due to the errors in the PPN parameters $\gamma$ and $\beta$, currently known at the $\simeq 10^{-5}$  accuracy level \citep{2003Natur.425..374B,2014CeMDA.119..237P,2015CeMDA.123..325F,2017AJ....153..121P,2018NatCo...9..289G}.
Instead, the error in the Uranian gravitational parameter $\mu_{\uranus}$ would be of less concern. Indeed, according to \citet{2014AJ....148...76J}, its current relative uncertainty amounts to just
\eqi
\rp{\sigma_{\mu_{\uranus}}}{\mu_{\uranus}} = 7\times 10^{-7}.
\eqf
However, \rfrs{numoJ2}{numoGE} and the pictures in the middle row of Figure\,\ref{fig2} show that the issue of the 1pN bias on the determination of $J_2^{\uranus}$ is negligible; suffice it to say that the current relative uncertainty in $J_2^{\uranus}$ is $2\times 10^{-4}$, as per Table 12 of \citet{2014AJ....148...76J}.
\subsubsection{The impact of the uncertainty in the pole of Uranus}\lb{polo}
Here, the consequences of the current level of uncertainty in our knowledge of the orientation of the Uranian spin axis on the otherwise idealized situation outlined in Section\,\ref{sec4} is examined.

At present, the pole of Uranus is known to an accuracy of \citep{2014AJ....148...76J}
\begin{align}
\sigma_{\alpha_{\uranus}} \lb{erRA} & = 0.002^\circ = 7.2\,\mathrm{arcsec}, \\ \nonumber \\
\sigma_{\delta_{\uranus}} \lb{erDEC} & = 0.002^\circ = 7.2\,\mathrm{arcsec}.
\end{align}
Such a somewhat modest accuracy is due to the present lack of dedicated, in-situ missions, contrary to the case, of, \textcolor{black}{e.g.}, Cassini \textcolor{black}{at} Saturn \citep{2019Sci...364.1046S}; for the Kronian spin axis, \citet{2022AJ....164..199J} recently obtained
\begin{align}
\sigma_{\alpha_{\saturn}} & \simeq 0.0001^\circ = 0.36\,\mathrm{arcsec}, \\ \nonumber\\
\sigma_{\delta_{\saturn}} & \simeq 0.00003^\circ = 0.1\,\mathrm{arcsec}.
\end{align}

The mismodelled quadrupole-induced orbital precessions due to the uncertainty in the primary's spin axis can be analytically evaluated as
\eqi
\sigma_{\dot\kappa^{J_2}} = \sqrt{\ton{\derp{\dot \kappa^{J_2}}{\alpha}}^2\,\sigma_\alpha^2 + \ton{\derp{\dot \kappa^{J_2}}{\delta}}^2\,\sigma_\delta^2},\,\kappa= I,\,\Omega,\lb{dkp}
\eqf
starting from \rfrs{iobl}{Oobl}.

For the $2\,000\times 100\,000\,\mathrm{km}$ orbit of \textcolor{black}{\rfr{per1}}, \rfr{dkp}, applied to \rfrs{iobl}{Oobl} with \rfrs{et}{inkl} and \rfrs{erRA}{erDEC}, yields
\begin{align}
\sigma_{\dot I^{J_2}} & = 40\,502\,\mathrm{mas\,yr}^{-1}, \\ \nonumber \\
\sigma_{\dot\Omega^{J_2}} & = 10\,983\,\mathrm{mas\,yr}^{-1},
\end{align}
which are about 3 orders of magnitude larger than the nominal LT rates of \rfrs{numILT}{numOLT}.

In the case of the less eccentric $2\,000\times 10\,000\,\mathrm{km}$ orbit of \textcolor{black}{\rfr{per2}}, one has
\begin{align}
\sigma_{\dot I^{J_2}} & = 324\,220\,\mathrm{mas\,yr}^{-1}, \\ \nonumber \\
\sigma_{\dot\Omega^{J_2}} & = 87\,918\,\mathrm{mas\,yr}^{-1},
\end{align}
which are $\simeq 2-3$ orders of magnitude larger than \rfrs{kazILT}{kazOLT}.

Thus, $\alpha_{\uranus}\,\delta_{\uranus}$ should be determined with an error as little as
\eqi
\sigma_{\alpha_{\uranus}}\simeq\sigma_{\delta_{\uranus}}\simeq 0.1-1\,\mathrm{mas}.\lb{errpole}
\eqf

A \textcolor{black}{further potential challenge is}  represented by the uncertainty $\sigma_I$ in the orbital inclination $I$ to the CE; \textcolor{black}{applying \rfr{dkp} to such a potentially} major source of systematic \textcolor{black}{bias yields that $I$ would need to be determined with an accuracy} as good as
\eqi
\sigma_I\simeq 1-10\,\mathrm{mas}.\lb{errI}
\eqf
\subsection{The range-rate and rate shifts}\lb{randrr}
One of the most accurately measured observable in a spacecraft-based mission to a planet P is the Earth-probe range-rate $\dot\rho$. As an example, the two-way Ka-band Doppler range-rate measurements of the spacecraft Juno \citep{juno18}, currently orbiting Jupiter, are accurate to a $\simeq 1\times 10^{-2}$ millimetre per second (mm s$^{-1}$) level over $60$ s after having processed data covering up to the middle of its prime mission \citep{2020GeoRL..4786572D}.

For the \textcolor{black}{computational strategy to recover the shift} $\Delta\dot\rho$ of the dynamical part of the range-rate due to the planetocentric orbital motion affected by a perturbing acceleration, see Appendix \ref{appenB}.

In the case of the LT effect, the gravitomagnetic acceleration ${\bds A}_\mathrm{LT}$  can be found, e.g., in \citet{Sof89,iers10,SoffelHan19}, and its radial, transverse and out-of-plane components in terms of the unit vectors $\bds{\hat{l}},\,\bds{\hat{m}},\,\bds{\hat{h}}$ are explicitly displayed in \citet{2017EPJC...77..439I}.
The resulting relevant gravitomagnetic shifts, calculated for the particular orbital geometry of \rfrs{et}{inkl}, turn out to be
\begin{align}
\Delta a^\mathrm{LT}\ton{f} & = 0,\\ \nonumber \\
\Delta e^\mathrm{LT}\ton{f} & = 0, \\ \nonumber \\
\Delta I^\mathrm{LT}\ton{f} \nonumber \lb{palle1} &= \rp{G\,S}{2\,c^2\,\nk\,a^3\ton{1-e^2}^{3/2}}\,\grf{
4\,\ton{f - f_0}\,\cos\delta +
 4\,\cos\ton{f + f_0 - \delta + 2\omega}\,\sin\ton{f - f_0} + \right.\\ \nonumber \\
\nonumber &+\left.  e\,\qua{3\,\sin\ton{f - \delta} - 3\,\sin\ton{f_0 - \delta} +\,\sin\ton{f + \delta} - \sin\ton{f_0 + \delta} +\,\sin\ton{f - \delta + 2\omega} + \right.\right.\\ \nonumber \\
&+\left.\left.   \sin\ton{3 f - \delta + 2\omega} - 2\,\cos f_0\,\sin\ton{2 f_0 - \delta + 2\omega}}
}, \\ \nonumber \\
\Delta\Omega^\mathrm{LT}\ton{f} \nonumber \lb{palle2} &= -\rp{G\,S}{2\,c^2\,\nk\,a^3\ton{1-e^2}^{3/2}}\,\grf{
2\,\cos\ton{2 f - \delta + 2\omega} -
 2\,\cos\ton{2 f_0 - \delta + 2\omega} + \right.\\ \nonumber \\
\nonumber &+\left.  e\,\qua{-3\,\cos\ton{f - \delta} + 3\,\cos\ton{f_0 - \delta} + \cos\ton{f + \delta} -
   \cos\ton{f_0 + \delta} + \cos\ton{f - \delta + 2\omega} + \right.\right.\\ \nonumber \\
 &+\left.\left.  \cos\ton{3 f - \delta + 2\omega} - 2\,\cos f_0\,\cos\ton{2 f_0 - \delta + 2\omega}} + 4\,\ton{-f + f_0}\,\sin\delta
}, \\ \nonumber \\
\Delta\omega^\mathrm{LT}\ton{f} & =0, \\ \nonumber \\
\Delta\mathcal{M}^\mathrm{LT}\ton{f} & = 0,
\end{align}
where $f$ is the true anomaly reckoning the instantaneous position of the test particle along its orbit, $f_0$ is the value of the true anomaly at some initial instant $t_0$, and $\mathcal{M}$ is the mean anomaly.
Thus, for the LT perturbations of the radial ($R$), transverse ($T$) and out-of-plane ($N$) components of the velocity one has
\begin{align}
\Delta v_R^\mathrm{LT}\ton{f} \lb{One}& = 0, \\ \nonumber \\
\Delta v_T^\mathrm{LT}\ton{f} & = 0, \\ \nonumber \\
\Delta v_N^\mathrm{LT}\ton{f} \lb{Three} \nonumber & = \rp{G\,S}{2\,c^2\,a^2\,\ton{1-e^2}^2}\,\ton{
4\,e\,\ton{f - f_0}\,\cos\ton{\delta - \omega} + 4\,\ton{f - f_0}\,\cos\ton{f - \delta + \omega} + \right.\\ \nonumber \\
\nonumber &+\left. 4\,\cos\ton{f_0 - \delta + \omega}\,\sin\ton{f - f_0} +
e\,\grf{\sin\ton{f - 3 f_0 + \delta - \omega} + \sin\ton{f - f_0 + \delta - \omega} + \right.\right.\\ \nonumber \\
\nonumber & +\left.\left. 6\,\sin\ton{2 f - \delta + \omega} + \sin\ton{f - f_0 - \delta + \omega} - 3\,\sin\ton{f + f_0 - \delta + \omega} - \right.\right.\\ \nonumber \\
\nonumber &-\left.\left. 2\,\sin\ton{2 f_0 - \delta + \omega} + e\,\qua{\sin\ton{f + \delta - \omega} - \sin\ton{f_0 + \delta - \omega} + \right.\right.\right.\\ \nonumber \\
&+ \left.\left.\left. 4\,\sin\ton{f - \delta + \omega} + \sin\ton{3 f - \delta + \omega} - 4\,\sin\ton{f_0 - \delta + \omega} - \sin\ton{3 f_0 - \delta + \omega}}}
}.
\end{align}
Inserting \rfrs{One}{Three} in \rfr{drr} allows to obtain the instantaneous LT range-rate shift;
\begin{align}
\Delta\dot\rho^\mathrm{LT}\ton{f} \lb{drrLT} \nonumber & = -\rp{G\,S\,\cos\chi\,\sin\ton{\alpha-\phi}}{2\,c^2\,a^2\,\ton{1-e^2}^2}\,\ton{
4\,e\,\ton{f - f_0}\,\cos\ton{\delta - \omega} + 4\,\ton{f - f_0}\,\cos\ton{f - \delta + \omega} + \right.\\ \nonumber \\
\nonumber &+\left. 4\,\cos\ton{f_0 - \delta + \omega}\,\sin\ton{f - f_0} +
e\,\grf{\sin\ton{f - 3 f_0 + \delta - \omega} + \sin\ton{f - f_0 + \delta - \omega} + \right.\right.\\ \nonumber \\
\nonumber & +\left.\left. 6\,\sin\ton{2 f - \delta + \omega} + \sin\ton{f - f_0 - \delta + \omega} - 3\,\sin\ton{f + f_0 - \delta + \omega} - \right.\right.\\ \nonumber \\
\nonumber &-\left.\left. 2\,\sin\ton{2 f_0 - \delta + \omega} + e\,\qua{\sin\ton{f + \delta - \omega} - \sin\ton{f_0 + \delta - \omega} + \right.\right.\right.\\ \nonumber \\
&+ \left.\left.\left. 4\,\sin\ton{f - \delta + \omega} + \sin\ton{3 f - \delta + \omega} - 4\,\sin\ton{f_0 - \delta + \omega} - \sin\ton{3 f_0 - \delta + \omega}}}
},
\end{align}
where $\phi,\,\chi$ are the RA and DEC of the planet, respectively.

After having averaged \rfr{drrLT} over one orbital revolution as
\eqi
\ang{\Delta\dot\rho^\mathrm{LT}} = \rp{\nk}{2\uppi}\,\int_{f_0}^{f_0+2\uppi}\Delta\dot\rho^\mathrm{LT}\ton{f}\,\dert{t}{f}\,\mathrm{d}f,
\eqf
\textcolor{black}{and considering} \rfr{dtdf},
one finally gets for the net gravitomagnetic range-rate shift per orbit
\eqi
\ang{\Delta\dot\rho^\mathrm{LT}} = -\rp{2\,G\,S\,\cos\chi\,\sin\ton{\alpha-\phi}\sin\ton{u_0 - \delta}}{c^2\,a^2\,\sqrt{1-e^2}\,\ton{1 + e\,\cos f_0}}.\lb{DNLT}
\eqf

The RA $\phi_{\uranus}$ and DEC $\chi_{\uranus}$ of Uranus, retrieved \textcolor{black}{through the web interface} HORIZONS maintained by the NASA Jet Propulsion Laboratory (JPL) for the epoch, say, 13 December 2032, are
\begin{align}
\phi_{\uranus} \lb{phiU} & = 89.455^\circ, \\ \nonumber\\
\chi_{\uranus} \lb{chiU}& = 23.6497^\circ.
\end{align}
\Rfr{DNLT}, calculated with \rfr{RA}, \textcolor{black}{\rfrs{Sura}{per1}} and \rfrs{phiU}{chiU}, yields
\eqi
\ang{\Delta\dot\rho^\mathrm{LT}}\,\qua{\rp{1 + e\,\cos f_0}{\sin\ton{u_0-\delta}}}= -\rp{2\,G\,S_{\uranus}\,\cos\chi_{\uranus}\,\sin\ton{\alpha_{\uranus}-\phi_{\uranus}}}{c^2\,a^2\,\sqrt{1-e^2}} =  8\times 10^{-5}\,\mathrm{mm\,s}^{-1}.\lb{gloz}
\eqf
By selecting
\begin{align}
f_0 \lb{guga}&= 180^\circ, \\ \nonumber\\
\omega \lb{condiz} & = \delta_{\uranus} - 90^\circ,
\end{align}
it is possible to maximize \rfr{gloz}\textcolor{black}{, thus getting}
\eqi
\ang{\Delta\dot\rho^\mathrm{LT}} = -\rp{2\,G\,S_{\uranus}\,\cos\chi_{\uranus}\,\sin\ton{\alpha_{\uranus} - \phi_{\uranus}}}{c^2\,a^2\,\sqrt{1-e^2}\,\ton{1-e}} = 2\times 10^{-4}\,\mathrm{mm\,s}^{-1}.\lb{gloz2}
\eqf

\begin{figure}[h]
\centering
\begin{tabular}{c}
\includegraphics[width = 16 cm]{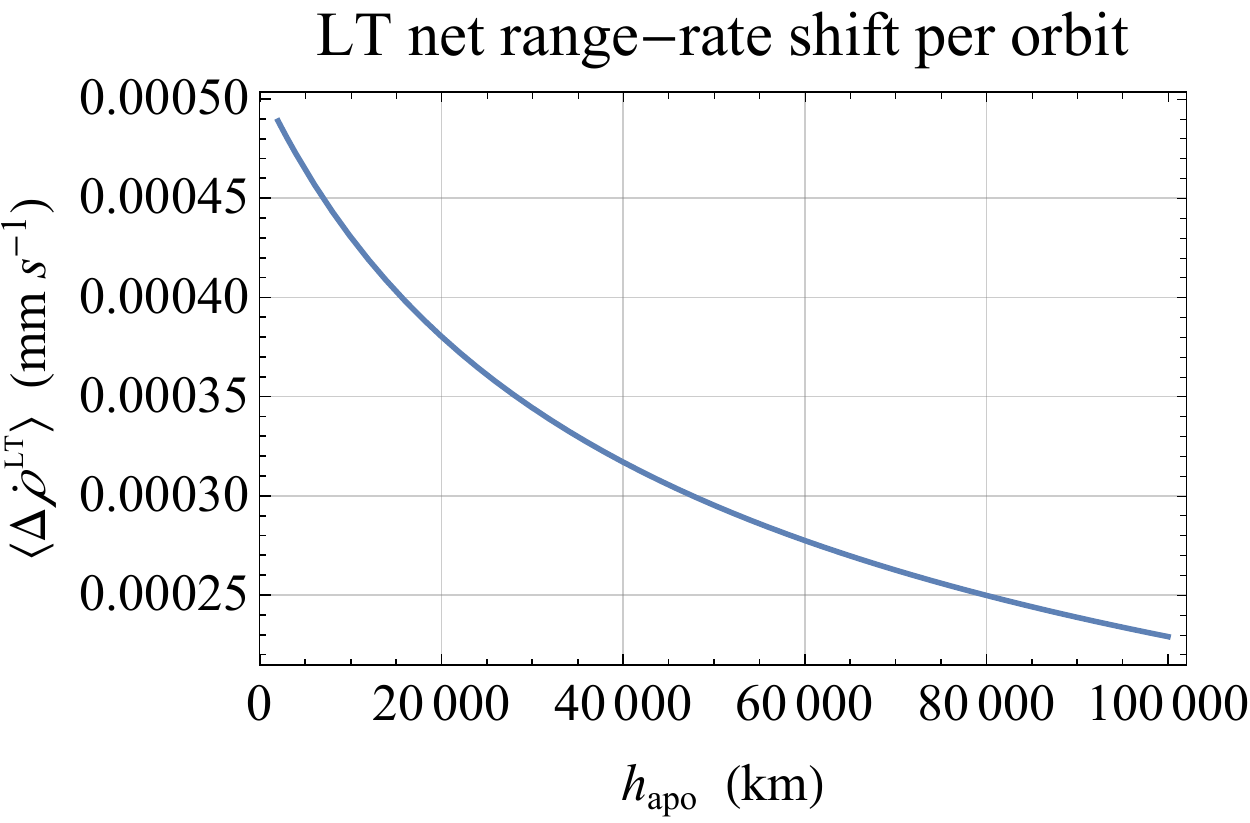}\\
\end{tabular}
\caption{
LT net shift per orbit $\ang{\Delta\dot\rho^\mathrm{LT}}$, in mm s$^{-1}$, of the range-rate of EURO calculated from \rfr{gloz2} as a function of  $h_\mathrm{apo}$, in km, for a fixed value of  $h_\mathrm{peri}=2\,000\,\mathrm{km}$. For $\alpha_{\uranus},\,S_{\uranus},\,\phi_{\uranus},\,\chi_{\uranus}$, \rfr{RA}, \rfr{Sura}, and \rfrs{phiU}{chiU} were adopted, respectively.
}\label{fig3}
\end{figure}

Figure\,\ref{fig3} plots \rfr{gloz2}, in mm s$^{-1}$, as a function of the apocenter height $h_\mathrm{apo}$, ranging from 2\,000 to $100\,000$ km, for a fixed value of the pericentre height $h_\mathrm{peri}=2\,000\,\mathrm{km}$.
\clearpage
It turns out that an apocentre height $h_\mathrm{apo}\simeq 50\,000\,\mathrm{km}$ allows to reach the $\simeq 3\times 10^{-4}\,\mathrm{mm\,s}^{-1}$ level; for $h_\mathrm{apo}=10\,000\,\mathrm{km}$, the LT net range-rate shift is of the order of $\simeq 4\times 10^{-4}\,\mathrm{mm\,s}^{-1}$.

\begin{figure}[h]
\centering
\begin{tabular}{c}
\includegraphics[width = 16 cm]{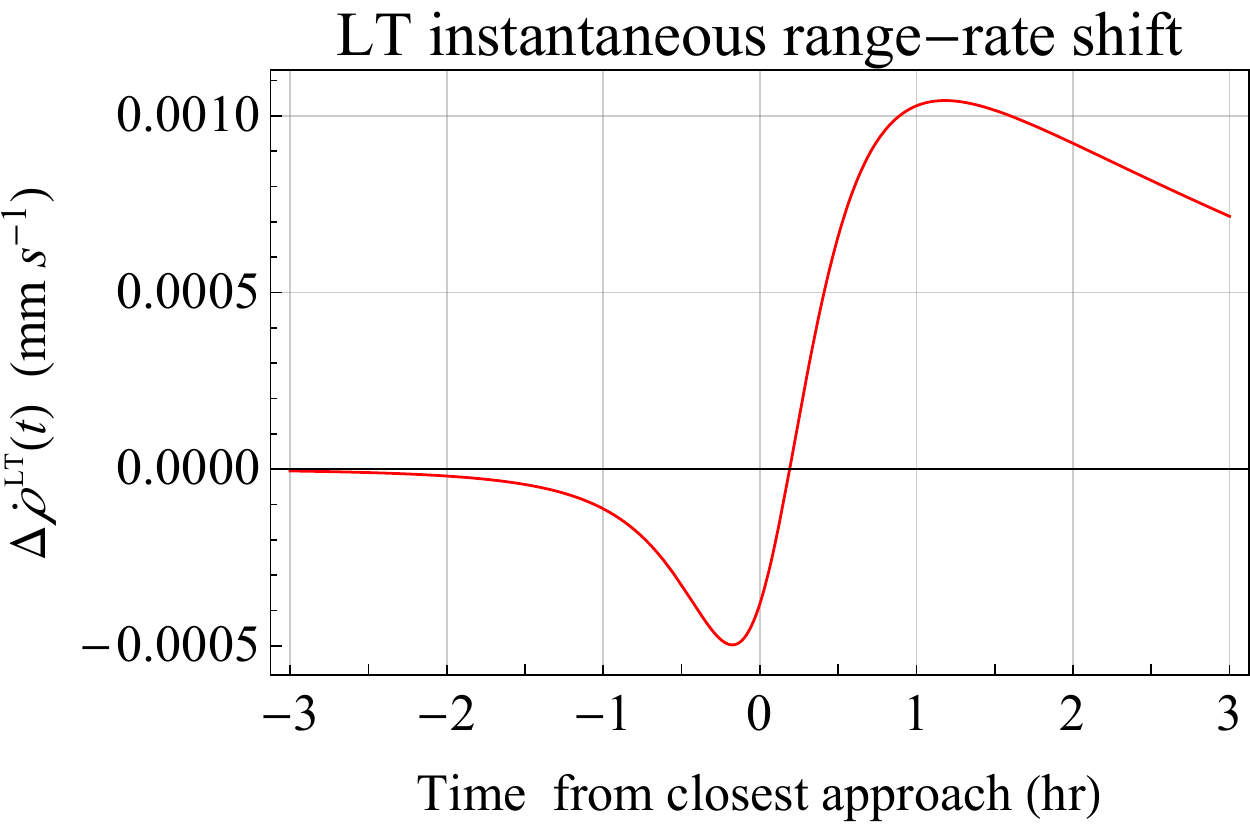}\\
\end{tabular}
\caption{
LT instantaneous shift $\Delta\dot\rho^\mathrm{LT}\ton{t}$, in mm s$^{-1}$, of the range-rate of EURO calculated from \rfr{drrLT} with \textcolor{black}{\rfr{per1}}, corresponding to a $2\,000\times 100\,000\,\mathrm{km}$ orbit, and \rfrs{guga}{condiz} as a function of time $t$, according to \rfr{ft} for $s_\mathrm{max} = j_\mathrm{max} = 55$,  during a passage 6 hours long centered at the pericentre. For $\alpha_{\uranus},\,\delta_{\uranus},\,S_{\uranus},\,\phi_{\uranus},\,\chi_{\uranus}$, \rfrs{RA}{DEC}, \rfr{Sura} and \rfrs{phiU}{chiU},  were adopted, respectively.
}\label{fig4}
\end{figure}

Figure\,\ref{fig4} depicts the instantaneous LT range-rate shift $\Delta\dot\rho^\mathrm{LT}\ton{t}$ during a single passage at the pericentre six hours long for the orbital configuration of \textcolor{black}{\rfr{per1}} and \rfrs{guga}{condiz}. To this end, we expressed $\Delta\dot\rho^\mathrm{LT}\ton{\textcolor{black}{f}}$  from \rfr{drrLT} as a function of time by switching from  $f$ to time $t$ through  $\mathcal{M}$ according to \citet{1961mcm..book.....B}
\eqi
f\ton{t} = \mathcal{M}\ton{t}
+
2\,\sum_{s=1}^{s_\mathrm{max}}\,\rp{1}{s}\,\grf{
\mathrm{J}_s\ton{s\,e} + \sum_{j=1}^{j_\mathrm{max}}\,\lambda^j\,\qua{\mathrm{J}_{s-j}\ton{s\,e} + \mathrm{J}_{s+j}\ton{s\,e}}}\,
\sin\ton{s\mathcal{M}\ton{t}},\lb{ft}
\eqf
where
\begin{align}
\mathcal{M}\ton{t} &=\mathcal{M}_0 +\nk\,\ton{t - t_0},\\ \nonumber \\
\lambda &= \rp{1-\sqrt{1-e^2}}{e},
\end{align}
$\mathcal{M}_0$ is the mean anomaly at the initial epoch $t_0$,
and $\mathrm{J}_k\ton{s\,e}$ is the Bessel function of the first kind of order $k$.
The initial instant $t_0$ corresponds to the passage at apocentre, as per \rfr{guga}, so that
\eqi
\mathcal{M}_0 = 180^\circ;
\eqf
the pericentre is reached after half a orbital period.
\clearpage
It can be noted that the peak-to-peak amplitude of the signal amounts to $\simeq 1.5\times 10^{-3}\,\mathrm{mm\, s}^{-1}$.

\textcolor{black}{Such figures are challenging if compared with the $\simeq 10^{-2}\,\mathrm{mm\,s}^{-1}$ accuracy level of Juno.}
\section{Orbital configurations and requirements}\lb{sec6}
\textcolor{black}{Elliptical orbits are generally preferred in mission design, since they allow to} greatly reduce the amount of propellant needed to impart the large change $\Delta V$ in speed required to finally insert the spacecraft in an circular, low trajectory \citep{2016PhDT.......195H}. Moreover, it is also mandatory to avoid that the more or less energetic particles circling in the radiation belts around the giant planets \citep{2011PEPI..187...92S,2022GeoRL..4900921M} damage the instrumentation of the spacecraft, which would likely happen if the latter steadily moved at the same (small) distance from the planet in a circular, low orbit. Another drawback of an (almost) \textcolor{black}{circular orbit is that the spacecraft} would go behind Uranus every few hours, making, thus, difficult to communicate with the Earth. Another issue is that low orbits must avoid the Uranian ring system. Should the probe move  completely inside the rings, the extended Uranian atmosphere \textcolor{black}{would cause its orbit to quickly decay. Moreover, there would}  exist a significant risk of frequent ring plane crossing.

Here, it is shown how the technique of aerocapture \citep{1979atfs.conf..195C,Lesz98,2019JSpRo..56..536S,2021JSpRo..58..505S,2022JSpRo..59.1074G} allows to insert a larger orbiter into an elliptical $2\,000\times 100\,000\,\mathrm{km}$ orbit, and a small secondary satellite into a closer and less elliptical $2\,000\times 10\,000\,\mathrm{km}$ orbit at Uranus; the orbital effects of interest for such orbital geometries were studied in Section\,\ref{sec5}. As shown by \rfr{per1} and \rfr{per2}, the resulting orbital periods are relatively short, if compared to Juno or Cassini Grand Finale orbits. Such a feature may be challenging for spacecraft operations, and orbital maneuvers should be carefully planned.
%
%

\begin{figure}[h]
\centering
\begin{tabular}{c}
\includegraphics[width =16 cm]{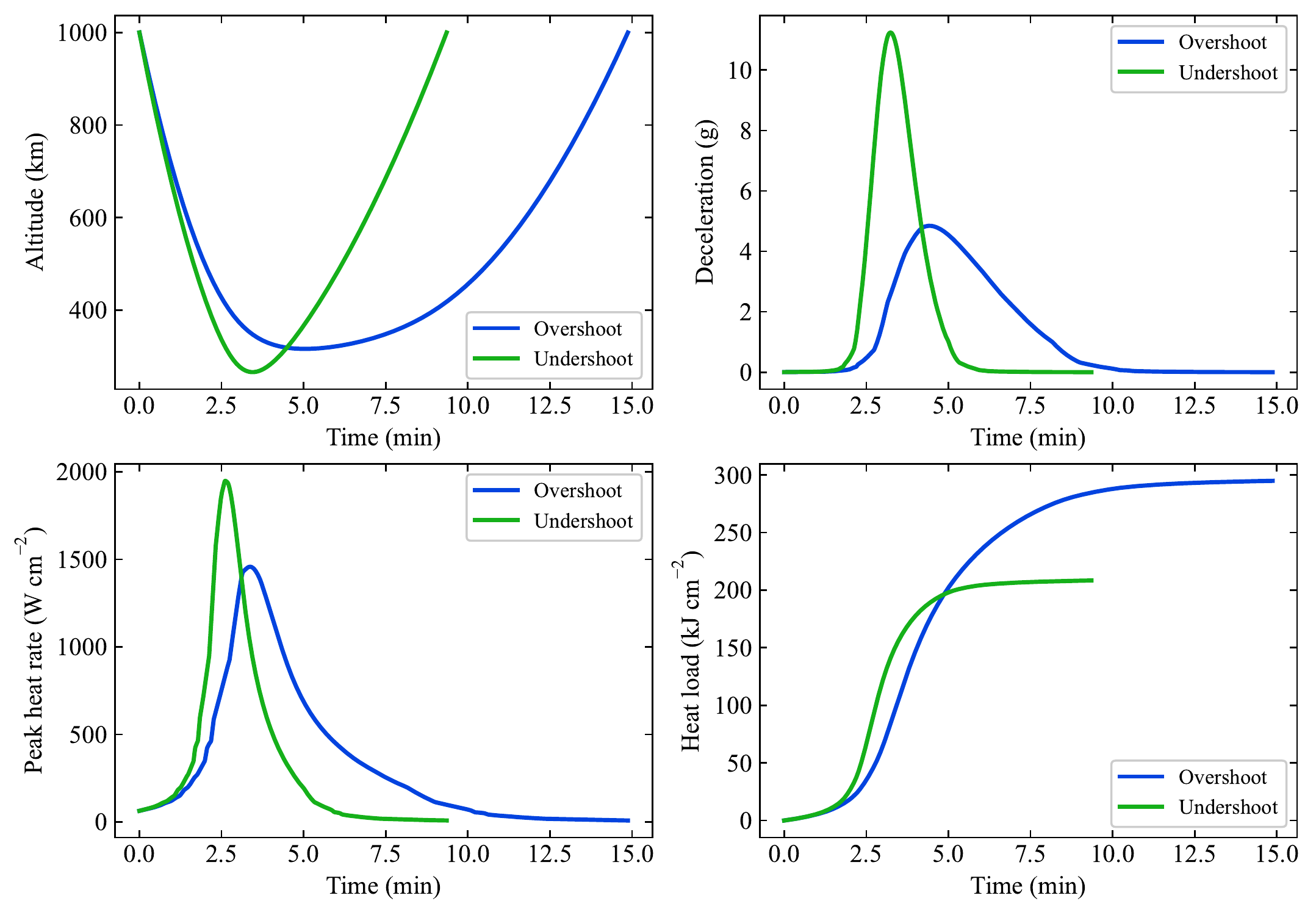}\\
\end{tabular}
\caption{
Overshoot and undershoot trajectories for a Mars Science Laboratory (MSL)-derived aerocapture vehicle with lift-to-drag $L/D = 0.24$, ballistic coefficient $\upbeta = 146\,\mathrm{kg\,m}^{-2}$, spacecraft nose radius $R_n = 1.125\,\mathrm{m},\  h_\mathrm{apo}= 100\,000\,\mathrm{km}$. The independent variable on the horizontal axes is the time since entry interface, defined at 1\,000 km above the 1 bar pressure level, while the unit of measure for the deceleration denoted as \virg{g} in the vertical axis of the panel in the upper right corner is the mean Earth's acceleration of gravity.
}\label{fig5}
\end{figure}

The mission and aerocapture vehicle design presented in \citet{2023AcAau.202..104G} to achieve a $4\,000\times 550\,000\,\mathrm{km}$ trajectory is
adapted here to achieve a smaller $2\,000\times 100\,000\,\mathrm{km}$ elliptical, polar orbit.
Its apocentre is just outside the Uranian ring system. Such a relatively small orbit is beyond the reach of conventional propulsive insertion due to the large $\Delta V$ required, but it turns out to be feasible using aerocapture. Using a vehicle with lift-to-drag $\ton{L/D} = 0.24$, the Theoretical Corridor Width (TCW) for aerocapture is $1.22^\circ$, which is sufficient to accommodate delivery errors and atmospheric uncertainties.
Figure\,\ref{fig5} shows the overshoot and undershoot limit trajectories for an aerocapture vehicle targeting a $100\,000\,\mathrm{km}$ apocentre
altitude at atmospheric exit using the Aerocapture Mission Analysis Tool by \citet{2021JOSS....6.3710G}.
\clearpage
The peak heat rate is estimated to be in the range of $1\,500$ to $2\,000\,\mathrm{W\,cm}^{-2}$, which is well within the capability of the Heatshield for Extreme Entry Environment Technology (HEEET) \citep{2020SSRv..216...22V}. At the first apocentre, the spacecraft performs a $120\,\mathrm{m\,s}^{-1}$ propulsive burn to raise the pericentre outside the atmosphere and achieve its $2\,000\times 100\,000\,\mathrm{km}$ elliptical capture orbit.
%
%

\begin{figure}[h]
\centering
\begin{tabular}{c}
\includegraphics[width =16 cm]{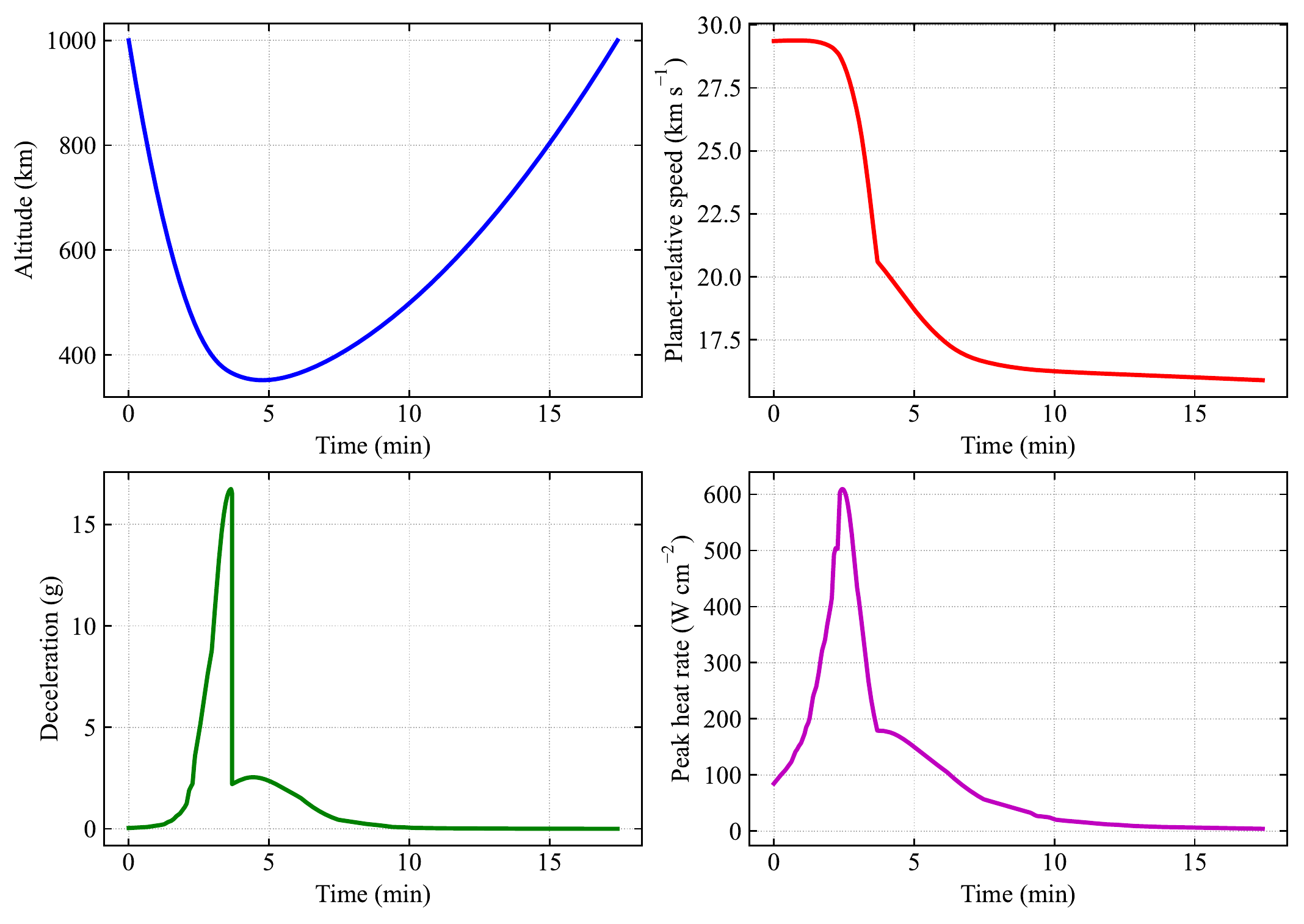}\\
\end{tabular}
\caption{
Nominal drag modulation trajectory for a vehicle with ballistic coefficient before the drag skirt separation$\upbeta_1 = 10\,\mathrm{kg\,m}^{-2}$, ballistic coefficient ratio $\upbeta_2/\upbeta_1 = 7.5$, and  apocentre altitude $h_\mathrm{apo}= 10\,000\,\mathrm{km}$.  The independent variable on the horizontal axes is the time since entry interface, defined at 1\,000 km above the 1 bar pressure level, while the unit of measure for the deceleration denoted as \virg{g} in the vertical axis of the panel in the lower left corner is the mean Earth's acceleration of gravity.
}\label{fig6}
\end{figure}

Using a different aerocapture vehicle design known as drag modulation, it is possible to send a small orbiter with, say,  a mass of $200\,\mathrm{kg}$,
into a $2\,000\times 10\,000\,\mathrm{km}$ as a secondary spacecraft carried by the main orbiter. The orbit is so small that the $\Delta V$ for
its insertion exceeds $12\,\mathrm{km\,s}^{-1}$. However, with aerocapture such an orbital configuration is achievable. It would also
be considered too risky for a large orbiter, and presents engineering challenges such as the spacecraft frequently going
behind Uranus as viewed from Earth and orbit decay due to drag from the upper atmosphere. A small spacecraft can
accommodate the risks while providing measurements from the low-circular orbit which cannot be made otherwise.
The small orbiter will detach from the mothership a few weeks prior to entry, and use aerocapture to insert into a
near-circular orbit around Uranus. The apocentre is just inside the rings, which start at $\simeq 12\,500\,\mathrm{km}$. Using
a drag modulation vehicle with\footnote{The parameters $\upbeta_1$ and $\upbeta_2$ are the values of the ballistic coefficient $\upbeta$ before and after the drag skirt separation, respectively, for drag modulation aerocapture. The ballistic coefficient ratio $\upbeta_2/\upbeta_1$ is a key design parameter indicating how much control authority the vehicle has.} $\upbeta_1 = 10\,\mathrm{kg\,m}^{-2}$ and $\upbeta_2/\upbeta_1 = 7.5$ offers a TCW of $0.70^\circ$ for aerocapture to a $h_\mathrm{apo} = 10\,000\,\mathrm{km}$. Figure\,\ref{fig6} shows the nominal aerocapture trajectory of the latter. At the first pericentre, the spacecraft performs a $200\,\mathrm{m\,s}^{-1}$ propulsive burn to achieve its initial capture orbit of $2\,000\times 10\,000\,\mathrm{km}$.
\clearpage
\section{Summary and conclusions}\lb{sec7}
A mission concept, provisionally dubbed EURO (Elliptical Uranian Relativity Orbiter), based on the use of a spacecraft around Uranus to measure the planet's angular momentum ${\bds S}_{\uranus}$ by means of its general relativistic LT orbital precessions independently of the Newtonian ones induced by the planetary gravity field's zonal multipoles $J_\ell,\,\ell=2,\,3,\,4,\ldots$ was presented.

The orbital geometry allowing, in principle, the fulfillment of such a goal implies \textcolor{black}{an orbital plane perpendicular to the CE and containing the Uranian spin axis ${\bds{\hat{k}}}_{\uranus}$. Moreover, its} node $\Omega$ is equal to the right ascension $\alpha_{\uranus}$ of ${\bds{\hat{k}}}_{\uranus}$. Indeed, in this case, the classical precessions of the orbital inclination $I$ and of $\Omega$ due to the even and odd zonals of the planet's gravity field ideally vanish, contrary to the gravitomagnetic ones. The situation is reversed for the precessions of the argument of pericentre $\omega$.

For a $2\,000\times 100\,000\,\mathrm{km}$ orbit, it turns out that the general relativistic rates of $I$ and $\Omega$ amount to $\simeq 16-59\, \mathrm{mas\,yr}^{-1}$, while the size of the Newtonian pericentre precession due to $J_2$ is as large as $\simeq 173^\circ\,\mathrm{yr}^{-1}$. An approximate evaluation of the LT net shift per orbit of the range-rate yields $\simeq 2 \times 10^{-4}\,\mathrm{mm\,s}^{-1}$ for a suitable choice of the initial conditions, with a peak-to-peak amplitude up to $\simeq 1.5 \times 10^{-3}\,\mathrm{mm\,s}^{-1}$ for a single passage at the pericentre few hours long\textcolor{black}{. Such figures are demanding with respect to the $\simeq 10^{-2}\,\mathrm{mm\,s}^{-1}$ accuracy level of, say, Juno, posing a challenge to the success of the mission}. By lowering the apocentre height down to $10\,000\,\mathrm{km}$, it is possible to increase the relativistic signatures to the level of
$\simeq 100-400\,\mathrm{mas\, yr}^{-1}$.

A major source of systematic bias is represented by the accuracies with which the orientation of ${\bds{\hat{k}}}_{\uranus}$ and $I$ should be determined. It turns out that they are of the order of $\simeq 0.1-1\,\mathrm{mas}$ and $\simeq 1-10\,\mathrm{mas}$, respectively; the pole of Uranus is currently known with an accuracy of $\simeq 7\,\mathrm{arcsec}$ due to the lack of dedicated, in-situ mission(s). \textcolor{black}{To this aim, it is conceivable that a possible  \virg{phase 1} of the EURO mission may allow for good enough measurements of ${\bds{\hat{k}}}_{\uranus}$ and $I$ by standard geodetic techniques.}
%
%
%

Using the LT effect to dynamically measure the spin angular momentum of Uranus is definitely a challenging task.
\section*{Acknowledgements}
One of us (L.I.) is grateful to  L. Iess and L. Petrov  for useful information. \textcolor{black}{All the authors thank the referee for her/his attentive and constructive remarks.}
\section*{Data availability}
Aerocapture trajectory results presented in the study can be reproduced using Jupiter Notebooks available at
https://amat.readthedocs.io/en/master/other-notebooks.html$\#$euro-uranus-orbiter. The study results were made
with AMAT v2.2.22. An archived version of AMAT v2.2.22 is available at Zenodo. DOI:10.5281/zenodo.7542714
\bibliography{Uranusbib}{}
\begin{appendices}
\section{The averaged rates of $I,\,\Omega,\,\omega$ due to some even and odd zonal harmonics}\lb{appenA}
\renewcommand{\theequation}{A.\arabic{equation}}
\setcounter{equation}{0}
The averaged rates of the inclination $I$, of the longitude of the ascending node $\Omega$, and of the argument of pericentre $\omega$ induced by the zonal harmonics $J_\ell$ of degree $\ell=2,\,3,\,4,\ldots$ of the multipolar expansion of the classical part of the gravitational potential of a non-spherical body
can be computed, e.g., with the  planetary equations in the form of Lagrange \citep{Bertotti03}
\begin{align}
\dot I^{J_\ell}  & = \rp{1}{\nk\,a^2\,\sqrt{1 - e^2}\sin I}\,\ton{
  \derp{\ang{\Delta U_\ell}}\Omega - \cos I\,\derp{\ang{\Delta U_\ell}}\omega}, \\ \nonumber \\
\dot\Omega^{J_\ell} & = -\rp{1}{\nk\,a^2\,\sqrt{1 - e^2}\sin I}\,
  \derp{\ang{\Delta U_\ell}} I, \\ \nonumber \\
\dot\omega^{J_\ell} & = \rp{\cot I}{\nk\,a^2\,\sqrt{1 - e^2}}\,\derp{\ang{\Delta U_\ell}}{I} -\rp{\sqrt{1 - e^2}}{\nk\,a^2\,e}\,\derp{\ang{\Delta U_\ell}}{e},
\end{align}
where
\eqi
\ang{U_\ell} = \rp{\nk}{2\uppi}\,\int_{f_0}^{f_0+2\uppi}\,\Delta U_\ell\,\dert{t}{f}\,\mathrm{d}f.\lb{Ul}
\eqf
In \rfr{Ul}, the instantaneous correction $\Delta U_\ell$ of degree $\ell$ of the Newtonian monopole, to be evaluated onto an unperturbed Keplerian ellipse by means of
\begin{align}
r &= \rp{p}{1 + e\,\cos f},\lb{rKep} \\ \nonumber \\
\dert{t}{f} &= \rp{\ton{1-e^2}^{3/2}}{\nk\,\ton{1+e\,\cos f}^2},\lb{dtdf}
\end{align}
is \citep{Bertotti03}
\eqi
\Delta U_\ell = \rp{\mu}{r}\,J_\ell\,\ton{\rp{R}{r}}^\ell\,\mathcal{P}_\ell\ton{\bds{\hat{k}}\bds\cdot\bds{\hat{r}}}\lb{DUl}.
\eqf
In \rfr{DUl}, $\mathcal{P}_\ell\ton{\cdots}$ is the Legendre polynomial of degree $\ell$, and its argument, i.e. the cosine of the angle between the primary's spin axis $\bds{\hat{k}}$ and the position vector $\bds r$ of the test particle, can be written as \citep{Sof89,1991ercm.book.....B,SoffelHan19}
\eqi
\bds{\hat{k}}\bds\cdot\bds{\hat{r}} = \ton{\bds{\hat{k}}\bds\cdot\bds{\hat{l}}}\cos u + \ton{\bds{\hat{k}}\bds\cdot\bds{\hat{m}}}\sin u,
\eqf
where
\eqi
u \doteq \omega + f
\eqf
is the argument of latitude.

For $\ell = 2,\,3,\,4,\,\ldots 8$, it turns out that the averaged rates of $I$ and $\Omega$, computed with \rfrs{et}{inkl}, vanish, contrary to the pericentre whose  precessions are
\begin{align}
\dot\omega^{J_2} \lb{oj2}& = -\rp{3}{4}\,\nk\,J_2\,\ton{\rp{R}{p}}^2, \\ \nonumber \\
e\,\dot\omega^{J_3} \lb{oj3}& = -\rp{3}{8}\,\nk\,J_3\,\ton{\rp{R}{p}}^3\,\ton{1+4\,e^2}\,\cos\xi,\\ \nonumber \\
\dot\omega^{J_4} \lb{oj4}& = -\rp{15}{128}\,\nk\,J_4\,\ton{\rp{R}{p}}^4\,\qua{
12 + 9\,e^2 + 2\,\ton{2 + 5\,e^2}\,\cos 2\xi
},\\ \nonumber \\
e\,\dot\omega^{J_5} \lb{oj5}& = -\rp{15}{256}\,\nk\,J_5\,\ton{\rp{R}{p}}^5\,\cos\xi\,\qua{8 + 75\,e^2 + 22\,e^4 + 14\,\ton{e^2 + 2\,e^4}\,\cos 2\xi}, \\ \nonumber \\
\dot\omega^{J_6} \nonumber \lb{oj6}& = -\rp{105}{4,096}\,\nk\,J_6\,\ton{\rp{R}{p}}^6\,\qua{
80 + 50\,e^2\,\ton{4 + e^2} + 10\,\ton{4 + 22 e^2 + 7 e^4}\,\cos 2\xi + \right.\\ \nonumber \\
&\left. + 3\,e^2\,\ton{4 + 7\,e^2}\,\cos 4\xi
},\\ \nonumber \\
e\,\dot\omega^{J_7} \nonumber \lb{oj7}& = -\rp{21}{8,192}\,\nk\,J_7\,\ton{\rp{R}{p}}^7\,\cos\xi\,\qua{
200 + 3,360\,e^2 + 3,570\,e^4 + 592\,e^6 + \right.\\ \nonumber \\
&\left. +  6\,e^2\,\ton{180 + 685\,e^2 + 136\,e^4}\,\cos 2\xi +
 33\,e^4\,\ton{5 + 8\,e^2}\,\cos 4\xi
}, \\ \nonumber \\
\dot\omega^{J_8} \nonumber \lb{oj8}& = \rp{63}{262,144}\,\nk\,J_8\,\ton{\rp{R}{p}}^8\,\ton{
-175\,\grf{64 + 7 e^2\,\qua{48 + 5\,e^2\,\ton{8 + e^2}}} - \right.\\ \nonumber \\
\nonumber &\left. - 70\,\ton{96 + 944\,e^2 + 970\,e^4 + 135\,e^6}\,\cos 2\xi -
 154\,e^2\,\ton{40 + 128\,e^2 + 27\,e^4}\,\cos 4\xi -\right.\\ \nonumber \\
&\left. -  286\,e^4\,\ton{2 + 3\,e^2}\,\cos 6\xi
},
\end{align}
where
\eqi
\xi\doteq \delta-\omega.
\eqf
Note that the precessions induced by the odd zonals, i.e. \rfr{oj3},\,\rfr{oj5}, and \rfr{oj7}, vanish if calculated with \rfr{condiz}.
\section{The range-rate perturbation}\lb{appenB}
\renewcommand{\theequation}{B.\arabic{equation}}
\setcounter{equation}{0}
The velocity vector $\bds v$ of a test particle moving along an unperturbed Keplerian ellipse around a massive primary  is
\eqi
\bds v = \grf{v_x,\,v_y,\,v_z},
\eqf
with \citep{2011rcms.book.....K,2014grav.book.....P}
\begin{align}
v_x \lb{vx} & = -\rp{\nk\,a}{\sqrt{1-e^2}}\,\qua{\cos\Omega\,\ton{e\,\sin\omega + \sin u} + \cos I\,\sin\Omega\,\ton{e\,\cos\omega + \cos u}}, \\ \nonumber \\
v_y \lb{vy} & = -\rp{\nk\,a}{\sqrt{1-e^2}}\,\qua{-\cos I\,\cos\Omega\,\ton{e\,\cos\omega + \cos u} + \sin\Omega\,\ton{e\,\sin\omega + \sin u}}, \\ \nonumber \\
v_z \lb{vz}& = \rp{\nk\,a}{\sqrt{1-e^2}}\,\sin I\,\ton{e\,\cos\omega + \cos u}.
\end{align}

The instantaneous change $\Delta\bds v$ of $\bds v$ has, then, to be computed as\footnote{In calculating the partial derivatives with respect to $a$, $\nk$ has to be considered as a function of $a$.}
\eqi
\Delta\bds v =\grf{\sum_{\kappa}\,\derp{v_x}{\kappa}\,\Delta\kappa\ton{f},\,\sum_{\kappa}\,\derp{v_y}{\kappa}\,\Delta\kappa\ton{f},\,\sum_{\kappa}\,\derp{v_z}{\kappa}\,\Delta\kappa\ton{f}},\,\kappa=a,\,e,\,I,\,\Omega,\,\omega,\,f.\lb{Deltavu}
\eqf
In \rfr{Deltavu}, the  instantaneous shift $\Delta\kappa\ton{f}$ of any perturbed osculating orbital element $\kappa$ among $a,\,e,\,I,\,\Omega,\,\omega$ can be analytically worked out as
\eqi
\Delta\kappa\ton{f} = \int_{f_0}^f\dert{\kappa}{t}\,\dert{t}{f^{'}}\,\mathrm{d}f^{'},\lb{Dk}
\eqf
where  $\mathrm{d}\kappa/\mathrm{d}t$ is the right-hand-side of the equation for the osculating element $\kappa$ in the Euler-Gauss form \citep{Sof89,1991ercm.book.....B,2011rcms.book.....K,SoffelHan19}, calculated with the disturbing acceleration $\bds A$ at hand and evaluated onto the unperturbed Keplerian ellipse characterized by \rfrs{rKep}{dtdf}.
As per the variation $\Delta f$ of the true anomaly, it is usally expressed in terms of the changes of the mean anomaly $\mathcal{M}$ and of the eccentricity $e$ as follows.
First, the true anomaly $f$ is written as a function of the eccentric anomaly $E$ as
\citep{Capde05}
\eqi
\tan\ton{\rp{f}{2}} = \sqrt{\rp{1+e}{1-e}}\,\tan\ton{\rp{E}{2}}\lb{fE}.
\eqf
Then, the shift of $f$  is calculated by differentiating \rfr{fE} with respect to $e$ and $E$ as
\eqi
\Delta f = \derp{f}{e}\,\Delta e +\derp{f}{E}\,\Delta E = \rp{\sin E}{\sqrt{1-e^2}\,\ton{1-e\,\cos E}}\,\Delta e + \sqrt{\rp{1+e}{1-e}}\,\rp{1-e}{\ton{1-e\,\cos E}}\,\Delta E\lb{Df}
\eqf
In turn, from the Kepler equation \citep{Capde05}
\eqi
\mathcal{M} = E -e\,\sin E,
\eqf
one gets
\eqi
\Delta E = \rp{\Delta\mathcal{M}+\sin E\,\Delta e}{1-e\,\cos E}\lb{DE}.
\eqf
Finally, by inserting \rfr{DE} in \rfr{Df} and using \citep{Capde05}
\begin{align}
r &= a\,\ton{1-e\,\cos E}, \\ \nonumber \\
\sin E & = \rp{\sqrt{1-e^2}\,\sin f}{1 + e\,\cos f}, \\ \nonumber \\
\cos E & =\rp{e + \cos f}{1 + e\,\cos f},
\end{align}
one obtains
\eqi
\Delta f = \ton{\rp{a}{r}}\,\qua{\sin f\,\ton{1 + \rp{r}{p}}\,\Delta e +\sqrt{1-e^2}\,\ton{\rp{a}{r}}}\,\Delta\mathcal{M},\lb{Deltaf}
\eqf
in agreement with \citet[Eq.\,(A.6)]{1993CeMDA..55..209C}.

The instantaneous perturbations $\Delta v_R\ton{f},\,\Delta v_T\ton{f},\,\Delta v_N\ton{f}$ of the radial ($R$), transverse ($T$) and out-of-plane ($N$) components of the velocity are worked out by projecting \rfr{Deltavu} onto the radial, transverse and normal directions given by the unit vectors \citep{Sof89,1991ercm.book.....B,SoffelHan19}
\begin{align}
{\bds u}_R \lb{ur} &= \grf{\cos\Omega\,\cos u - \cos I\,\sin\Omega\,\sin u,\, \sin\Omega\,\cos u + \cos I\,\cos\Omega\,\sin u,\, \sin I\,\sin u}, \\ \nonumber \\
{\bds u}_T \lb{ut} &= \grf{-\cos\Omega\,\sin u - \cos I\,\sin\Omega\,\cos u,\, -\sin\Omega\,\sin u + \cos I\,\cos\Omega\,\cos u,\, \sin I\,\cos u}, \\ \nonumber \\
{\bds u}_N \lb{un} &= \grf{\sin I\,\sin\Omega,\, -\sin I\,\cos\Omega,\, \cos I}.
\end{align}

As a result, one gets
\begin{align}
\Delta v_R\ton{f} \lb{dvr} \nonumber & = \Delta\bds v\bds\cdot{\bds u}_R =  -\rp{\nk\,\sin f}{\sqrt{1 - e^2}}\,\qua{\rp{e\,\Delta a\ton{f}}{2} + \rp{a^2\,\Delta e\ton{f}}{r}} - \\ \nonumber \\
&- \rp{\nk\, a^2\,\sqrt{1 - e^2}}{r}\,\qua{\cos I\,\Delta\Omega\ton{f} + \Delta\omega\ton{f}} - \rp{\nk\,a^3}{r^2}\,\Delta\mathcal{M}\ton{f}, \\ \nonumber \\
\Delta v_T\ton{f} \lb{dvt} \nonumber &= \Delta\bds v\bds\cdot{\bds u}_T =  -\rp{\nk\,a\,\sqrt{1 - e^2}}{2\, r}\,\Delta a\ton{f} + \rp{\nk\, a\,\ton{e + \cos f}}{\ton{1 - e^2}^{3/2}}\,\Delta e\ton{f} + \\ \nonumber \\
& + \rp{\nk\, a\, e\,\sin f}{\sqrt{1 - e^2}}\,\qua{\cos I\,\Delta\Omega\ton{f} + \Delta\omega\ton{f}}, \\ \nonumber \\
\Delta v_N\ton{f} \lb{dvn} &= \Delta\bds v\bds\cdot{\bds u}_N = \rp{\nk\,a}{\sqrt{1-e^2}}\,\qua{\ton{\cos u + e\,\cos\omega}\,\Delta I\ton{f} + \sin I\,\ton{\sin u + e\,\sin\omega}\,\Delta\Omega\ton{f}},
\end{align}
which agree just  with \citet[Eqs.\,(33)-(35)]{1993CeMDA..55..209C}.
Thus, the perturbation of the velocity vector can be expressed as
\eqi
\Delta\bds v = \Delta v_R\,{\bds{u}}_R + \Delta v_T\,{\bds{u}}_T + \Delta v_N\,{\bds{u}}_N\lb{Deltav},
\eqf
where $\Delta v_R,\,\Delta v_T,\,\Delta v_N$ are given by \rfrs{dvr}{dvn}, and the unit vectors ${\bds{u}}_R,\,{\bds{u}}_T,\,{\bds{u}}_N$ are as per \rfrs{ur}{un}.

In case of a spacecraft orbiting a distant planet P, the dynamical\footnote{It means that it is solely due to the planetocentric orbital motion of the probe; it neglects all the special and general relativistic effects connected with the propagation of the electromagnetic waves through a variable gravitational field \citep{2011rcms.book.....K}.} component of the range-rate $\dot\rho$ is the projection of the planetocentric velocity $\bds v$ onto the unit vector $\bds{\hat{\rho}}$ of the line of sight which can  be approximated with the opposite of the versor of the geocentric position vector $\bds{\hat{\mathrm{R}}}$ of P. Usually, the latter remains constant during a fast passage of the probe at the pericentre, or even during a full orbital revolution. In terms of the RA $\phi$ and DEC $\chi$ of P, it can be written as
\eqi
\bds{\hat{\rho}}\simeq -\bds{\hat{\mathrm{R}}} = -\grf{\cos\chi\,\cos\phi,\,\cos\chi\,\sin\phi,\,\sin\chi}.\lb{rho}
\eqf
Thus, the range-rate perturbation $\Delta\dot\rho$ can be analytically calculated by means of \rfr{Deltav} and \rfr{rho} as
\eqi
\Delta\dot\rho = \Delta\bds v\bds\cdot\bds{\hat{\rho}} \simeq -\Delta\bds v\bds\cdot\bds{\hat{\mathrm{R}}}.\lb{drr}
\eqf
\section{Naming the apsidal positions in the case of Uranus}\lb{appenC}
\renewcommand{\theequation}{C.\arabic{equation}}
\setcounter{equation}{0}
Here, some proposals for properly naming the pericentre and apocentre in the case of Uranus are given.

The first suggestion is \virg{periouranon} and \virg{apouranon}, from
\grk{per\'{i}}$ + $\grk{O>uran\'{o}s} (\textrm{per\'{\i}}$ + $\textrm{\={U}ran\'{o}s}) and \grk{>ap\'{o}}$ + $\grk{O>uran\'{o}s} (\textrm{ap\'{o}}$ + $\textrm{\={U}ran\'{o}s}), respectively. Indeed, \grk{O>uran\'{o}s} is the Greek god personifying the sky after whom the seventh planet of the solar system was named, while \grk{per\'{i}} ($+$ accusative) and \grk{>ap\'{o}} ($+$ genitive) are prepositions meaning \textit{around}, \textit{near}, \textit{about}, and \textit{from}, \textit{away from}, respectively. One might be tempted to propose the term \virg{aphouranon}, as in \virg{aphelion} for the Sun. Actually, the former should be avoided since the apocopic form \grk{>af'} (aph$^\prime$) of \grk{>ap\'{o}} is used before a wovel with rough breathing \grk{(<)}, as just in \grk{<\'{H}lios} (H$\acute{\bar{\mathrm{e}}}$lios), while \grk{O>uran\'{o}s} has the smooth breathing \grk{(>)}.

As an alternative proposal, \virg{peric{\ae}lum} and \virg{apoc{\ae}lum} could also be considered  since \textit{C{\ae}l$\breve{u}$s} is  the Roman counterpart of \grk{O>uran\'{o}s}.

Should one rely upon the colour of the planet, \virg{peric{\ae}rulum} and \virg{apoc{\ae}rulum} could be adopted, from the Latin adjective \textit{c{\ae}r$\breve{u}$lus} meaning, among other things, just \virg{greenish-blue}.
\end{appendices}
\end{document}